\documentclass[reprint,
superscriptaddress,
amsmath,amssymb,
aps,
prx,
]{revtex4-2}
\usepackage{graphicx}% Include figure files
\usepackage{dcolumn}% Align table columns on decimal point
\usepackage{bm}% bold math
\usepackage{dsfont}
\usepackage{physics}

\usepackage{graphicx}% Include figure files
\usepackage{dcolumn}% Align table columns on decimal point
\usepackage{bm}% bold math
\usepackage[utf8]{inputenc}
\usepackage[T1]{fontenc}
\usepackage{mathptmx}
\usepackage{gensymb}
\usepackage[dvipsnames]{xcolor}
\usepackage{amssymb}
\usepackage{physics}
\usepackage{verbatim}
\usepackage{appendix}

\DeclareFontFamily{U}{wncy}{}
\DeclareFontShape{U}{wncy}{m}{n}{<->wncyr10}{}
\DeclareSymbolFont{mcy}{U}{wncy}{m}{n}
\DeclareMathSymbol{\Sh}{\mathord}{mcy}{"58} 

\usepackage{xr}

\begin{document}
\title{Reciprocal Asymptotically Decoupled Hamiltonian for Cavity Quantum Electrodynamics}

\author{Michael A.D. Taylor}
\email{michael.taylor@rochester.edu}
\affiliation{The Institute of Optics, Hajim School of Engineering, University of Rochester, Rochester, New York 14627, USA}

\author{Braden M. Weight}
\email{bweight@ur.rochester.edu}
\affiliation{Department of Physics and Astronomy, University of Rochester, Rochester, New York 14627, USA}

\author{Pengfei Huo}
\email{pengfei.huo@rochester.edu}
\affiliation{Department of Chemistry, University of Rochester, 120 Trustee Road, Rochester, New York 14627, USA} 
\affiliation{The Institute of Optics, Hajim School of Engineering, University of Rochester, Rochester, New York 14627, USA}

\date{\today}% It is always \today, today.

\begin{abstract}
We develop a new theoretical framework for describing light-matter interactions in cavity quantum electrodynamics (QED), optimized for efficient convergence at arbitrarily strong coupling strengths and is naturally applicable to low-dimensional materials. This new Hamiltonian is obtained by applying a unitary gauge transformation on the p$\cdot$A Hamiltonian, with a shift on both the matter coordinate and the photonic coordinate, then performing a phase rotation and transforming in the reciprocal space of the matter. By formulating the light-matter interaction in terms of an upper-bounded effective coupling parameter, this method allows one to easily converge eigenspectra calculations for any coupling strength, even far into the ultra-strong and deep-strong coupling regimes. We refer to this new approach as the Reciprocal Asymptotically Decoupled (RAD) Hamiltonian. The RAD Hamiltonian allows for a fast convergence of the polariton eigenspectrum with a much smaller matter and photon basis, compared to the commonly used p$\cdot$A or dipole gauge Hamiltonians. The RAD Hamiltonian also allows one to go beyond the commonly used long-wavelength approximation and accurately describes the spatial variations of the field inside the cavity, which ensures the conservation of momentum between light and matter. 
\end{abstract}

\maketitle

\section{Introduction}
Quantum electrodynamics has been extremely successful in describing the fundamental quantum interaction between light and matter~\cite{CohenTannoudji1997}. Different applications of this theory, from quantum optics~\cite{Jaynes1963PotI,Mischuck,Hofheinz2008,Hofheinz2009} to polariton chemistry~\cite{Flick2017PNAS,Ebbesen2016ACR,Feist2018,Ribeiro2018,Mandal2023CR}, have been at the forefront of physics. Many approximations, including the two-level approximation, the rotating-wave approximation, and the neglecting of second-order terms such as the dipole self-energy (in the multipolar gauge) or the diamagnetic term (in the Coulomb gauge) have historically been sufficient to replicate experimental results. However, in recent years, experimental advances in optical cavity design have produced light-matter coupling strengths, for which these approximations are no longer valid~\cite{Bernardis2018PRA,Bernardis2018PRAa,Kockum2019NRP,Mandal2023CR}. This, in conjunction with the recent increase in computational power, has led to a revival of exact, fundamental forms of cavity QED~\cite{Stefano2019NP,Bernardis2018PRA, Taylor2020PRL,Stokes2022RMP}.

The most fundamental cavity quantum electrodynamics (QED) Hamiltonian~\cite{CohenTannoudji1997} is the minimal coupling Hamiltonian (also known as the ``p$\cdot$A'' Hamiltonian). However, as widely discussed in the literature \cite{Bernardis2018PRAa,Taylor2020PRL,Stokes2022RMP,Stefano2019NP,Li2020PRB}, the p$\cdot$A Hamiltonian converges very slowly in terms of matter states~\cite{Bernardis2018PRAa, Li2020PRB}. One can resolve this issue caused by the truncation of this Hamiltonian's Hilbert space by carefully considering the proper way to apply the appropriate projection operators during the derivation of various Hamiltonians \cite{Taylor2020PRL,Taylor2022OL,Stefano2019NP}. Many others instead use the Pauli-Fierz Hamiltonian~\cite{Flick2017PNAS,Schaefer2018PRA,Rokaj2018JPBAMOP,Li2020PRB,haugland_QEDIntermol_JCP2021,haugland_QEDCC_PRX2020,riso_QEDHF_NatCommun2022,riso_QEDIONIZATION_JCP2022,weight_QEDPerespective_PCCP2023,vu_enhanced_JPCA2022,deprince_QEDIONIZATION_JCP2021,liebenthal_QEDEOMCCSD_JCP2022,mctague_nhQEDCIS_JCP2022,weight_abQED_JPCL2023,weight_diffusion_Arxiv2023,cui_vQEDMP2_arXiv2023,li_vQED_arXiv2023,vu_QEDCASCI_arXiv2023}, which is related to the minimal coupling Hamiltonian via the Power-Zienau-Woolley unitary transformation (see Appendix III), transforming it into the dipole gauge (multipolar gauge). The PF Hamiltonian, while requiring fewer matter states (compared to the ``p$\cdot$A'' Hamiltonian) has been shown to require significantly more Fock states to converge when the light-matter coupling strength is large.~\cite{Mandal2020JPCL,Li2020PRB}
These above-mentioned convergence difficulties for both common gauges become especially salient in the ultra-strong coupling regime~\cite{Kuisma2022AP,Canales2023N} and the deep-strong coupling regime~\cite{Kockum2019NRP}, where the Rabi splitting becomes greater than the cavity transition frequency\cite{Ashida2021PRL,Kockum2019NRP}. In this regime, even simple models become extraordinarily difficult to calculate in either the dipole or Coulomb gauge~\cite{Ashida2021PRL}. Although these regimes may seem unreachable experimentally, experimentalists in recent years have demonstrated such couplings on multiple occasions~\cite{Bayer2017NL,Yoshihara2016NP,Mueller2020N}. These experimental innovations require a new theoretical framework to accurately model them.
  
In a recent work~\cite{Ashida2021PRL}, Ashida {\it et al.} introduced a new representation to model cavity QED systems for arbitrary coupling strengths that use an effective coupling parameter, which has a global maximum at a finite value of the original coupling strength and then decays; hence, it is referred to as the Asymptotically Decoupled (AD) Hamiltonian. However, the AD Hamiltonian mediates light-matter coupling by shifting the matter coordinates in their external potential by a photonic operator. Unless the potential is of a specific form (such as a simple cosine function), even model-system calculations require applying a Taylor series expansion of the the potential. Taking inspiration from the AD Hamiltonian~\cite{Ashida2021PRL}, we introduce the Reciprocal Asymptotically Decoupled (RAD) Hamiltonian, which possesses the benefits of the AD Hamiltonian, but all components of it are separable between operators in the photonic and electronic DOFs, regardless of the potential. Expressed in Fourier space, the RAD Hamiltonian can be directly applied to periodic systems.

In formulating this novel representation of QED, we arrive at three key results that make this formalism widely applicable to many systems. We derive a general expression for the RAD Hamiltonian for many modes and particles presented in Eq.~\ref{eq:RAD_Ham}, which is the first {\it key result}. In the special case of a single particle coupled to a single cavity mode, the specific expression of the RAD Hamiltonian is expressed in Eq.~\ref{eq:h_rad-1par-1mode}. Then, we focus on the special case of periodic systems, where we can parameterize the Hamiltonian in terms of the lattice momentum $\bf k$, yielding $\hat{H}_\mathrm{RAD}({\bf k})$ as the second {\it key result} of this paper in Eq.~\ref{eq:h_rad_k_gen}. Finally, we generalize the formalism, going beyond the commonly-used long wavelength approximation (LWA) to arrive at our final {\it key result} in Eq.~\ref{eq:h_rad_period_blwa_k}. We show that many forms of the LWA currently in use violate the conservation of momentum between the light and matter DOFs, with our formulation explicitly preserving it. With these key results, we hope that this RAD Hamiltonian will shed new light on investigating cavity QED systems, especially for periodic systems like 2D materials and systems that go beyond the ultrastrong coupling regime. 

\section{Asymptotically Decoupled Hamiltonian} \label{sec:AD}
We present a brief derivation of the Asymptotically Decoupled (AD) Hamiltonian~\cite{Ashida2021PRL} to provide the context for the rest of this article. We begin our derivation of the AD Hamiltonian with the most fundamental QED Hamiltonian \cite{CohenTannoudji1997}, the minimal coupling Hamiltonian (also known as the ''p$\cdot$A'' Hamiltonian) in the Coulomb gauge ($\nabla\cdot {\bf A}=0$) as follows
\begin{equation}\label{eqn:Hc}
\hat{H}_\mathrm{p\cdot A}=\sum_j \frac{1}{2m_j}(\hat{\bf p}_j-{z}_j \hat{\bf A})^2+\hat{V}(\hat{\bf x})+ \sum_\beta \hbar\omega_\beta \Big(\hat{a}^{\dagger}_\beta \hat{a}_\beta +\frac{1}{2}\Big),
\end{equation}
where $\beta$ iterates over all photonic modes (with wavevectors, $\bf k_\beta$, and polarizations, $\lambda$) and $j$ is the index of the $j_\mathrm{th}$ charged particle (including all electrons and nuclei), with the corresponding mass, $m_j$, charge, $z_j$, and $\hat{\bf p}_j=-i\hbar{\boldsymbol \nabla}_j$ is the canonical momentum operator. The quantized vector potential, $\hat{\bf A} = \sum_\beta {\bf A}_\beta (\hat{a}_\beta+\hat{a}^{\dagger}_\beta)= \sum_\beta  {A}_\beta (\hat{a}_\beta+\hat{a}^{\dagger}_\beta)\hat{\bf e}_\beta$, is purely transverse, defined by the Coulomb gauge under the long-wavelength approximation, where $\hat{a}^{\dagger}_\beta$ and $\hat{a}_\beta$ are the raising and lowering operators of the photonic DOF that satisfy $[\hat{a}_\beta,\hat{a}^{\dagger}_{\beta'}]=\delta_{\beta,\beta'}$ and $\hat{\bf e}_\beta$ is the polarization direction of the $\beta_\mathrm{th}$ quantized electric field. Furthermore, $\hat{H}_\mathrm{ph}= \sum_\beta \hbar\omega_\beta (\hat{a}^{\dagger}_\beta \hat{a}_\beta + \frac{1}{2})$
is the pure photonic Hamiltonian.

Following the procedure in Ref.~\citenum{Ashida2021PRL}, we derive the Asymptotically Decoupled (AD) Hamiltonian generalized for many charged particles and many photonic modes. We first rewrite Eq.~\ref{eqn:Hc} in its expanded form as
\begin{align}\label{eqn:demlerHc}
    \hat{H}_\mathrm{p \cdot A} =&~ \hat{H}_\mathrm{M}\otimes\hat{\mathds{1}}_\mathrm{ph} + \hat{\mathds{1}}_\mathrm{e}\otimes \sum_\beta \hbar \omega_\beta (\hat{a}_\beta^\dagger \hat{a}_\beta + \frac{1}{2})  \\
    &- \sum_{ j,\beta} \frac{z_j \hat{\bf p}_j \cdot {\bf A}_\beta}{m_j} (\hat{a}^\dagger_\beta + \hat{a}_\beta) 
    + \hat{\mathds{1}}_\mathrm{e}\otimes \sum_{j} \frac{z_j^2 |\hat{\bf A} |^2 }{2 m_j}, \nonumber
\end{align}
where $\hat{H}_\mathrm{M}$ is the pure matter Hamiltonian, and $\hat{\mathds{1}}_\mathrm{e}$ and $\hat{\mathds{1}}_\mathrm{ph}$ are the identity operators in the electronic and photonic subspaces, respectively. For simplicity, we will omit writing these two identity operators unless explicitly mentioned. 

We introduce a new mode-dependent coupling parameter,  
\begin{equation} \label{eq:g}
    \gamma_\beta = |\mathbf{A_\beta}| \sqrt{ \left( \frac{\omega_\beta}{\hbar} \right) \sum_j \frac{z_j^2}{m_j}},
\end{equation}
where we can re-express the $\hat{H}_\mathrm{p \cdot A}$ as
\begin{align}\label{eq:HPA-gc}
\hat{H}_\mathrm{p \cdot A}=&~ \hat{H}_\mathrm{M} + \sum_\beta \hbar \omega_\beta (\hat{a}_\beta^\dagger \hat{a}_\beta + \frac{1}{2}) - \sum_{ j,\beta} \frac{z_j \hat{\bf p}_j \cdot {\bf A}_\beta}{m_j} (\hat{a}^\dagger_\beta + \hat{a}_\beta) \nonumber \\
&+\sum_{\beta,\beta'} \frac{\hbar \gamma_\beta \gamma_{\beta'}}{2 \sqrt{\omega_\beta \omega_{\beta'}}} (\hat{a}^\dagger_\beta + \hat{a}_\beta) (\hat{a}^\dagger_{\beta'} + \hat{a}_{\beta'}) (\hat{\bf e}_\beta \cdot \hat{\bf e}_{\beta'}),
\end{align}
where we have explicitly expanded $|\hat{\bf A}|^2$. Note that the coupling strength $\gamma_\beta$ has a unit of frequency, and $\gamma_\beta/\omega_\beta$ can be used as the unitless coupling parameter to characterize the light-matter coupling strength. In the second line of Eq.~\ref{eq:HPA-gc}, there are now direct coupling terms between different modes. Equivalently, this can be written in terms of the mode's photonic momentum, $\hat{p}_\beta$, and coordinate, $\hat{q}_\beta$ as follows
\begin{align}\label{eq:HPA-qp}
\hat{H}_\mathrm{p \cdot A}=&~ \hat{H}_\mathrm{M} -  \sum_{ j,\beta} \frac{z_j \hat{\bf p}_j \cdot {\bf A}_\beta}{m_j} \sqrt{\frac{2 \omega_\beta}{\hbar}}\hat{q}_\beta  \\
&+ \sum_{\beta,\beta'} \frac{1}{2} \Big[\hat{p}_\beta^2 \delta_{\beta,\beta'} + \big(\omega_\beta^2 \delta_{\beta,\beta'} + 2 \gamma_\beta \gamma_{\beta'}(\hat{\bf e}_\beta \cdot \hat{\bf e}_{\beta'}) \big) \hat{q}_\beta \hat{q}_{\beta'} \Big], \nonumber
\end{align}
where $\hat{p}_\beta$ and  $\hat{q}_\beta$ are defined as,
\begin{subequations} \label{eq:qp_beta}
\begin{align}\label{Eq:q-beta}
\hat{q}_\beta = \sqrt{\frac{\hbar}{2\omega_\beta}}(\hat{a}^{\dagger}_\beta + \hat{a}_\beta)\\
\hat{p}_\beta = i\sqrt{\frac{\hbar\omega_\beta}{2}}(\hat{a}^{\dagger}_\beta - \hat{a}_\beta) \label{Eq:p-b}
\end{align}
\end{subequations}
We can then perform a normal mode analysis (See Appendix~\ref{app:normal_modes}) on the second line of Eq.~\ref{eq:HPA-qp} to generate a set of non-interacting modes, $\{\alpha\}$, with transformed frequencies, $\{\Omega_\alpha\}$ As an aside, note that in the single-mode limit, this reduces to a Bogoliubov transform (see Appendix~\ref{app:bogoliubov}). The Coulomb gauge Hamiltonian in Eq.~\ref{eq:HPA-qp} then becomes
\begin{align}\label{eq:shift_pA}
&\hat{H}_\mathrm{p \cdot A}= \hat{H}_\mathrm{M} -  \sum_{ j,\alpha} \frac{z_j \hat{\bf p}_j \cdot {\bf A}_\alpha}{m_j} \sqrt{\frac{2 \Omega_\alpha}{\hbar}}\hat{q}_\alpha  + \sum_\alpha \frac{1}{2} \Big(\hat{p}_\alpha^2 + \Omega_\alpha^2 \hat{q}_\alpha^2 \Big) \nonumber \\
&= \hat{H}_\mathrm{M} + \sum_\alpha \frac{1}{2} \Big(\hat{p}_\alpha^2 + \Omega_\alpha^2 (\hat{q}_\alpha - \sum_j \hat{\bf p}_j \cdot \boldsymbol{\xi}_{j,\alpha} )^2  - \Omega_\alpha^2 \big( \sum_j \hat{\bf p}_j \cdot \boldsymbol{\xi}_{j,\alpha} \big)^2 \Big)
\end{align}
where further details on the normal mode transformation are contained in Appendix~\ref{app:normal_modes}, and the coupling strength ${\boldsymbol{\xi}}_{ j,\alpha}$ is expressed as
\begin{equation} \label{eq:xi_j_alpha}
    {\boldsymbol{\xi}}_{ j,\alpha} = \sqrt{\frac{2}{\hbar}} \frac{z_j}{m_j \Omega_\alpha^{3/2}} {\bf A}_\alpha.
\end{equation}
Note that the values of both $\Omega_\alpha$ and ${\bf A}_\alpha$ must be found by the normal mode transformation and can be represented as linear combinations of $\{ \omega_\beta \}$ and $\{ {\bf A}_\beta \}$, respectively.

Additionally, the $\hat{\bf p}^2$ terms in Eq.~\ref{eq:shift_pA} can be grouped and thought of as an effective kinetic energy for each $j_\mathrm{th}$ particle. As such, we combine this term with the matter kinetic energy operator in $\hat{H}_\mathrm{M}$ and refer to it as $\hat{T}_\mathrm{AD}$,
\begin{align} \label{eq:T_AD}
    \hat{T}_\mathrm{AD} &= \sum_{j} \bigg[ \frac{\hat{\bf p}_{j}^2}{2m_j} - \sum_{i,\alpha} \frac{1}{\hbar \Omega_\alpha} (\hat{\bf p}_j \cdot {\bf A}_\alpha)(\hat{\bf p}_i \cdot {\bf A}_\alpha) \frac{z_i z_j}{m_i m_j}  \bigg] \\
    &= \sum_{j} \bigg[ \frac{\hat{\bf p}_{j}^2}{2m_j} - \sum_{i,\alpha} \frac{\Omega_\alpha^2}{2} ({\boldsymbol{\xi}}_{ j,\alpha} \cdot \hat{\bf p}_j) ({\boldsymbol{\xi}}_{ i,\alpha} \cdot \hat{\bf p}_i)  \bigg] \nonumber.
\end{align}
This can be thought of as a light-dressed matter kinetic energy. 

Recall that a coordinate shift operator, $\hat{U}_q = e^{-\frac{i}{\hbar}q_0 \hat{p}}$ displaces $\hat{q}$ by the amount $q_0$, such that $\hat{U}_q^\dagger \hat{O}(\hat{q}) \hat{U}_q = \hat{O}(\hat{q} + q_0)$. Based on Ref.~\citenum{Ashida2021PRL}, we introduce  a unitary transformation operator, which is a shift operator in both photonic and matter ``coordinates'' as follows
\begin{equation} \label{eq:uad}
    \hat{U}_\mathrm{AD} = \exp \bigg[ {-\frac{i}{\hbar} \sum_{ j,\alpha} {\boldsymbol{\xi}}_{ j,\alpha} \cdot \hat{\bf p}_j \hat{p}_\alpha} \bigg].
\end{equation}
The above ``double-shift'' operator removes the photonic coordinate shift in Eq.~\ref{eq:shift_pA} but simultaneously creates a new shift in all matter coordinates $\{\hat{\bf x}_j \}$ (see Eq.~\ref{eqn:Hc}). This is analogous to the Lee-Low-Pines transformation in condensed matter, which transforms electron-phonon couplings~\cite{Lee1953}.

The Asymptotically Decoupled (AD) framework of the QED Hamiltonian can be obtained by applying this ``double-shift'' operator to the p$\cdot$A Hamiltonian (Eq.~\ref{eq:shift_pA}) through $\hat{H}_\mathrm{AD} = \hat{U}_\mathrm{AD}^\dagger \hat{H}_\mathrm{p \cdot A} \hat{U}_\mathrm{AD}$, resulting in 
\begin{equation}\label{eqn:AD_ham}
    \hat{H}_\mathrm{AD} = \hat{T}_\mathrm{AD} + \hat{V}(\{ \hat{\bf x}_j + \sum_\alpha {\boldsymbol{\xi}}_{ j,\alpha} \hat{p}_\alpha \}) + \sum_\alpha \hbar \Omega_\alpha \Big(\hat{b}_\alpha^\dagger \hat{b}_\alpha + \frac{1}{2}\Big),
\end{equation}
where $\hat{b}_\alpha$ and $\hat{b}^{\dagger}_\alpha$ are the photonic annihilation and creation operators for the $\alpha_\mathrm{th}$ normal mode and are expressed as
\begin{subequations} \label{eq:b_a}
    \begin{align}
    \hat{b}_\alpha &= \sqrt{\frac{\Omega_\alpha}{2 \hbar}} \hat{q}_\alpha + i \sqrt{\frac{1}{2 \hbar \Omega_\alpha}} \hat{p}_\alpha \\
    \hat{b}_\alpha^\dagger &= \sqrt{\frac{\Omega_\alpha}{2 \hbar}} \hat{q}_\alpha - i \sqrt{\frac{1}{2 \hbar \Omega_\alpha}} \hat{p}_\alpha
    \end{align}
\end{subequations}
Note that Eq.~\ref{eqn:AD_ham}, under the single-particle limit, is the key result of Ref.~\citenum{Ashida2021PRL}, where $\Omega_\alpha$ becomes a renormalized frequency (of the Bogoliubov transform) in the single-mode limit (see Eq.~\ref{norm-freq} in Appendix~\ref{app:bogoliubov}).

For clarity, below we will consider the single-particle case in a one-dimensional potential such that $\hat{\bf e}\cdot\hat{\bf p} = \hat{p}$. This consideration simplifies Eq.~\ref{eqn:AD_ham} to
\begin{equation}\label{HAD}
    \hat{H}_\mathrm{AD} = \frac{\hat{p}^2}{2m_\mathrm{eff}}+\hat{V}({\hat{x}} + \sum_\alpha {\xi}_\alpha \hat{p}_\alpha) + \sum_\alpha \hbar \Omega_\alpha \Big(\hat{b}^\dagger_\alpha \hat{b}_\alpha + \frac{1}{2} \Big),
\end{equation}
where in the single-particle, one-dimensional case, $\hat{T}_\mathrm{AD}$ can be simplified by defining the effective mass as
\begin{equation}
    \frac{1}{m_\mathrm{eff}} = \frac{1}{m} + \sum_\alpha {\Omega_\alpha^2 \xi_\alpha^2}.
\end{equation}
This Hamiltonian has the advantage of an effective coupling parameter ${\xi}_\alpha$, which has an upper bound~\cite{Ashida2021PRL} at a finite value of the original coupling strength, such that for arbitrarily high light-matter coupling, the effective coupling parameter tends to zero and hence \textit{decouples} the light and matter DOFs which provides rapid convergence in the number of basis states for each subsystem when computing eigenenergies.~\cite{Ashida2021PRL} However, Eq.~\ref{HAD} is also {\it inconvenient} for numerical calculations due to the requirement of a shift of the matter position operator ${\hat{x}}$ by a photonic operator ${\xi_\alpha} \hat{p}_\alpha$ inside the potential $\hat{V}$, which has pure imaginary matrix elements in the Fock state representation (recall $\hat{p}_\alpha$ in Eq.~\ref{Eq:p-b} is the photonic momentum operator). Thus, it is, in general, inconvenient to evaluate $\hat{V}({\hat{x}}\otimes\hat{\mathds{1}}_\mathrm{ph} + \hat{\mathds{1}}_\mathrm{e}\otimes\sum_\alpha {\xi_\alpha} \hat{p}_\alpha)$, unless one uses special properties of $\hat{V}$ for a certain type of potential (such as trigonometric identities when $\hat{V}(\hat{x})$ is a trigonometric function~\cite{Ashida2021PRL}). Furthermore, for any potential that is not translationally invariant, one might need to expand it as $\hat{V}(\hat{x}+\sum_\alpha {\xi_\alpha} \hat{p}_\alpha)\approx \hat{V}(\hat{x})+\sum_{l=1}^{l_\mathrm{max}}\hat{V}^{(l)}(\sum_\alpha {\xi_\alpha} \hat{p}_\alpha)^{l}$, where $\hat{V}^{(l)}(\hat{x})=\partial^{l} \hat{V}/\partial \hat{x}^{l}$, and the results could be sensitive to the truncation of the series $l_\mathrm{max}$~\cite{Ashida2021PRL}. The scope and applicability of this form of the QED Hamiltonian (Eq.~\ref{HAD} or Eq.~\ref{eqn:AD_ham}) will be significantly expanded if this problem can be circumvented, for example, by further transforming the Hamiltonian into reciprocal space. This will be the focus of the current paper, as we discuss in the next section.

\section{Reciprocal Asymptotically Decoupled Hamiltonian} \label{sec:RAD}
To address the challenges in $\hat{H}_\mathrm{AD}$ as mentioned above, we present the Reciprocal Asymptotically Decoupled (RAD) Hamiltonian. To derive this new Hamiltonian, we first introduce a unitary operator that performs a $\pi/2$ rotation in phase space for all photonic modes such that $\hat{U}_{\pi/2}^\dagger f(\hat{p}_\alpha) \hat{U}_{\pi/2} = f(\hat{q}_\alpha)$, where $\hat{p}_\alpha$ and $\hat{q}_\alpha$ are defined in Eqs.~\ref{Eq:q-beta}-\ref{Eq:p-b}. Such an operator has the following form,
\begin{equation}\label{EQ:Unitary_p_to_q_rotation}
    \hat{U}_{\pi/2} = \exp \big( - i \frac{\pi}{2} \sum_\alpha \hat{b}_\alpha^\dagger \hat{b}_\alpha \big).
\end{equation}

Applying the above phase rotation operator to the AD Hamiltonian in Eq.~\ref{eqn:AD_ham} yields
\begin{equation}\label{phase-shift}
\hat{U}_{\pi/2}^\dagger \hat{H}_\mathrm{AD} \hat{U}_{\pi/2} = \hat{T}_\mathrm{AD} + \hat{V}(\{ \hat{\bf x}_j + \sum_\alpha {\boldsymbol{\xi}}_{ j,\alpha} \hat{q}_\alpha \}) + \sum_\alpha \hbar \Omega_\alpha (\hat{b}_\alpha^\dagger \hat{b}_\alpha + \frac{1}{2})
\end{equation}
We find that, instead of having the position operator $\hat{\bf x}_j$ shifted by a pure imaginary operator inside the potential, $\hat{\bf x}_j$ is now shifted by $\sum_\alpha {\boldsymbol{\xi}}_{ j,\alpha} \hat{q}_\alpha$, an operator with purely real matrix elements in the Fock basis.
This seems to be a trivial transform; however, by enforcing a phase rotation (swapping $\hat{p}_\alpha$ with $\hat{q}_\alpha$ in Eq.~\ref{phase-shift}), this Hamiltonian is now {\it purely real}. The Hamiltonian in Eq.~\ref{phase-shift} drastically simplifies our task later. 

% Using the representation of $\hat{K'} = \hat{p}/\hbar$ by inserting the resolution of identity (for the electronic subspace) $\int dK' |K'\rangle \langle K'|$ on both sides, we have the Reciprocal Asymptotically Decoupled (RAD) form of the QED Hamiltonian $\hat{H}_\mathrm{RAD}$ as follows
To represent our operators in reciprocal space, we consider the eigenstates of the momentum operator, $\ket{{\bf K}_j}$, where $\hat{\bf p}_j \ket{{\bf K}_j} = \hbar {\bf K}_j \ket{{\bf K}_j}$. Similarly, the matter identity can then be represented as
\begin{equation} \label{eq:iden_k}
    \mathds{1}_\mathrm{M} = \int \{ d {\bf K}_i \} \,\, \bigotimes_i \ketbra{{\bf K}_i}{{\bf K}_i},
\end{equation}
where we are integrating over ${\bf K}_i$ for all $i$ particles. By applying this form of unity to a matter operator, we can represent that operator in reciprocal space. For example, $\hat{\bf p}_j$ can be written as
\begin{align} \label{eq:p_j_i}
    \hat{\bf p}_j &= \int \{ d {\bf K}_i \} \,\, \bigotimes_{i<j} \ketbra{{\bf K}_i}{{\bf K}_i} \otimes \hbar {\bf K}_j \ketbra{{\bf K}_j}{{\bf K}_j} \otimes \bigotimes_{i>j} \ketbra{{\bf K}_i}{{\bf K}_i}  \nonumber \\ 
    &\equiv \int \{ d {\bf K}_i \} \,\, \hbar {\bf K}_j \bigotimes_{i}  \ketbra{{\bf K}_i}{{\bf K}_i},
\end{align}
where $j \in \{1, 2, ... i, ... N \}$ for $N$ particles. Since $\hat{T}_\mathrm{AD}$ is only a function of $\{ \hat{\bf p}_j \}$, it is purely diagonal in this reciprocal representation. By inserting identity into Eq.~\ref{eq:T_AD}, we get
\begin{align} \label{Eqn:T_RAD}
\hat{T}_\mathrm{RAD} =\int \{ d {\bf K}_i \} \,\, \sum_j &\bigg[ \frac{|\hbar {\bf K}_j|^2}{2 m_j} - \sum_{l,\alpha} \frac{\hbar^2 \Omega_\alpha^2}{2} ({\boldsymbol{\xi}}_{ j,\alpha} \cdot {\bf K}_j) ({\boldsymbol{\xi}}_{ l,\alpha} \cdot {\bf K}_l)  \bigg]\nonumber\\
&\bigotimes_i \ketbra{{\bf K}_i}{{\bf K}_i} ,
\end{align}
where $j,l \in \{1, 2, ... i, ... N \}$. In this manner, we diagonalize this dressed kinetic energy term.

Similarly, the interaction term, $\hat{V}(\{ \hat{\bf x}_j + \sum_\alpha {\boldsymbol{\xi}}_{ j,\alpha} \hat{q}_\alpha \})$, can be expressed in the reciprocal space as follows
\begin{align}
\hat{V}_\mathrm{RAD} = \int & \{ d {\bf K}_i \}\{ d {\bf K}_i'' \}\{ d {\bf x}_i \}  ~ \, V(\{ {\bf x}_i + \sum_\alpha {\boldsymbol{\xi}}_{ i,\alpha} \hat{q}_\alpha \}) \\
& \bigotimes_i \ket{{\bf K}_i} \braket{{\bf K}_i}{{\bf x}_i} \braket{{\bf x}_i}{{\bf K}_i''} \bra{{\bf K}_i''}. \nonumber 
\end{align}
By using the identity, $\braket{{\bf K}_i}{{\bf x}_i} = \exp (-i {\bf K}_i \cdot {\bf x}_i) / \sqrt{2 \pi}$, we can then simplify our expression of $\hat{V}_\mathrm{RAD}$ to
\begin{align} \label{Eq:V_RAD_shift}
    \hat{V}_\mathrm{RAD} =&~ \int \{ d {\bf K}_i \}\{ d {\bf K}_i'' \}\{ d {\bf x}_i \}  ~ \, V(\{ {\bf x}_i + \sum_\alpha {\boldsymbol{\xi}}_{ i,\alpha} \hat{q}_\alpha \})  \\
    &~\times\prod_j \Big[ \frac{1}{2 \pi}  e^{i ({\bf K}_j'' - {\bf K}_j) \cdot {\bf x}_j} \Big] \bigotimes_i  \ketbra{{\bf K}_i}{{\bf K}_i''} \nonumber,
\end{align}
where $V(\{ {\bf x}_i + \sum_\alpha {\boldsymbol{\xi}}_{ i,\alpha} \hat{q}_\alpha \}) = \bra{{\bf x}_i} \hat{V}(\{ \hat{\bf x}_j + \sum_\alpha {\boldsymbol{\xi}}_{ j,\alpha} \hat{q}_\alpha \}) \ket{{\bf x}_i}$. The term inside the square brackets of Eq.~\ref{Eq:V_RAD_shift} is the Fourier kernel for this many-dimensional space, with the Fourier transform defined as 
\begin{equation}\label{eq:FT}
    \mathcal{F} \{ g(x) \} = \frac{1}{2 \pi} \int dx \, g(x) e^{i x K}
\end{equation}
This integral over $\{ {\bf x}_i \}$ in Eq.~\ref{Eq:V_RAD_shift} is now just a Fourier transform of $V(\{{\bf x}_i + \sum_\alpha {\boldsymbol{\xi}}_{ i,\alpha} \hat{q}_\alpha \})$ for all matter DOFs. 

\begin{figure}
 \centering
 %%-------start of first figure----------
    \begin{minipage}[h]{\linewidth}
     \centering
     \includegraphics[width=1.0\linewidth]{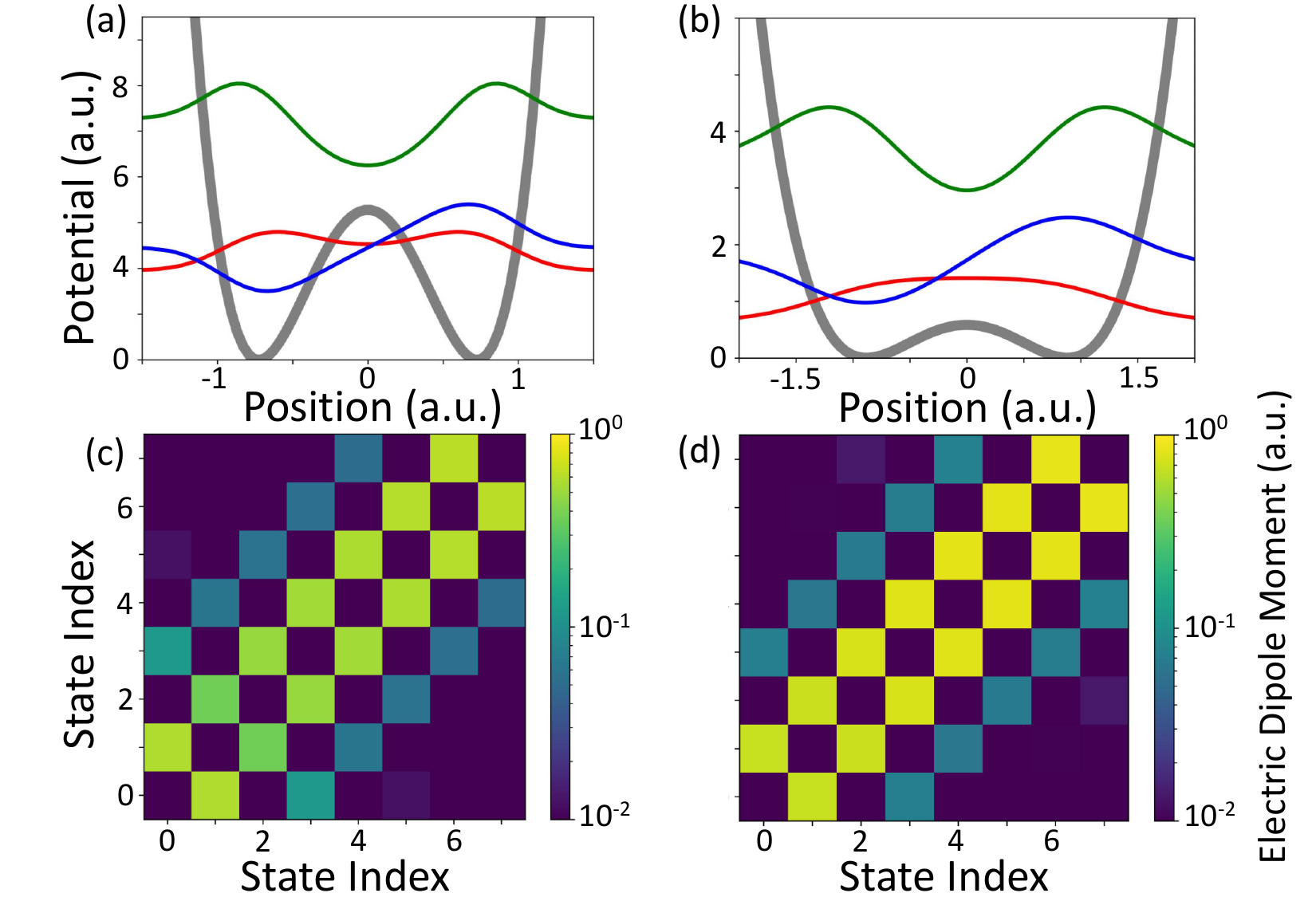}
    \end{minipage}%
   \caption{The potential $V(x)$ of the (a) steep and (b) shallow double-well models, along with the three lowest-energy wavefunctions for the electron confined in the potential. The dipole matrix of the (c) steep and (d) shallow double well potential are shown, which are responsible for mediating the interaction in the PF Hamiltonian (Eq.~\ref{eq:PFham}).}
   \label{FIG:MAIN_POTS_WFNS_DIPS}
\end{figure}

\begin{figure}
 \centering
 %%-------start of first figure----------
    \begin{minipage}[h]{\linewidth}
     \centering
     \includegraphics[width=0.9\linewidth,trim=0 0 0 0,clip]{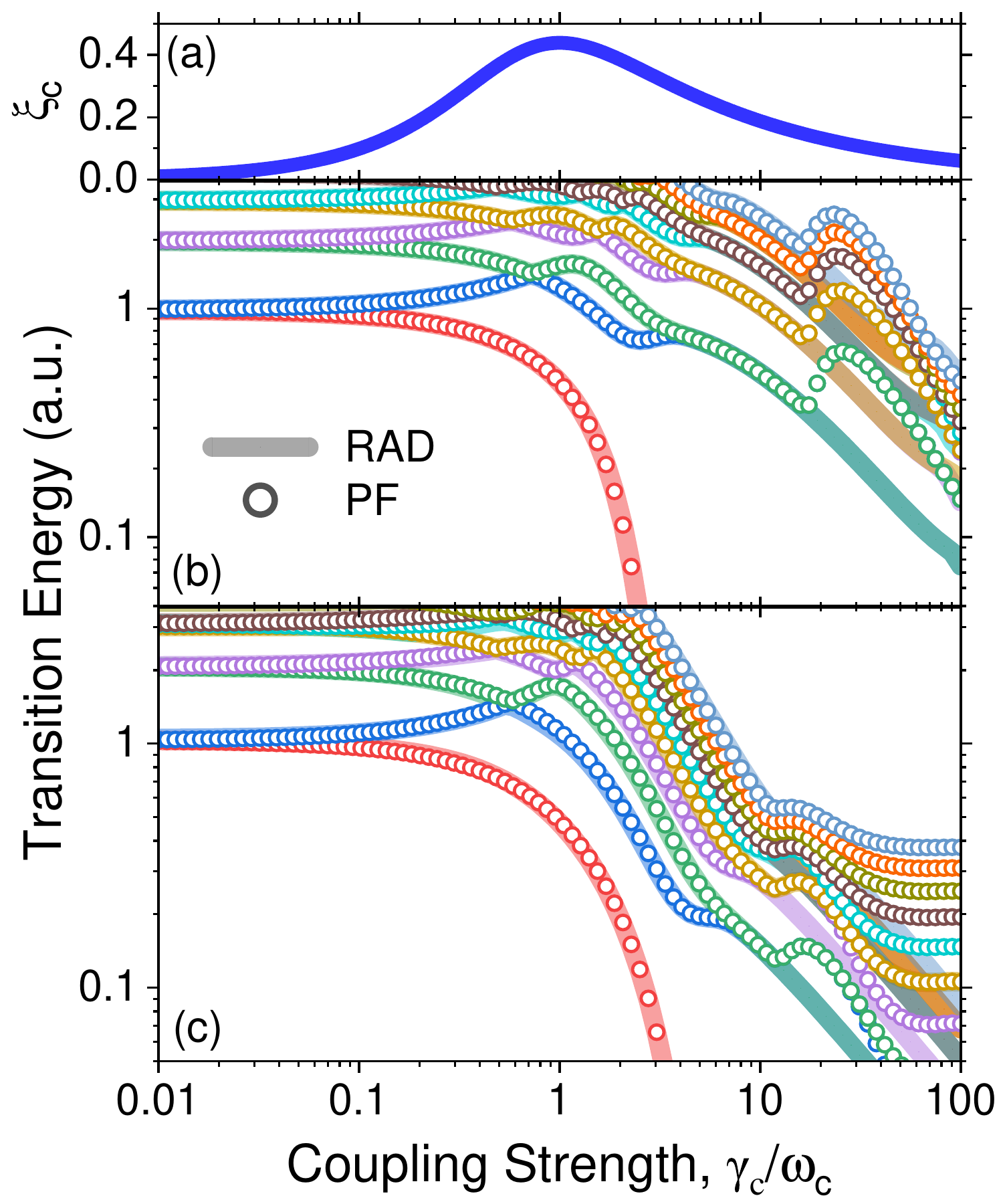}
    \end{minipage}%
   \caption{ (a) The effective coupling parameter $\xi_g$ (Eq.~\ref{eq:xi_j_alpha}) as a function of the coupling strength $\gamma_\mathrm{c}/\omega_\mathrm{c}$, which exhibits a bounded nature in RAD. Eigenspectrum of the (b) steep double-well potential (see Fig.~\ref{FIG:MAIN_POTS_WFNS_DIPS}a) and (c) shallow double-well potential (see Fig.~\ref{FIG:MAIN_POTS_WFNS_DIPS}b), obtained from the RAD Hamiltonian (solid lines) and the PF Hamiltonian (open circles). In both panels, the RAD Hamiltonian was diagonalized with 100 matter $K$-grid points and 20 Fock states while the PF Hamiltonian was diagonalized with 50 matter adiabatic states and 200 Fock states. The matter adiabatic states were obtained by diagonalizing $\hat{H}_\mathrm{M}$ with 2048 matter \textit{grid points} in real space to ensure that the adiabatic energies and dipoles are properly converged, before using them for constructing the PF Hamiltonian and diagonalization.}
   \label{FIG:MAIN_Xi_gc__loglog_Demler}
\end{figure}

Since the shift in the potential in Eq.~\ref{Eq:V_RAD_shift} is now real (after applying the unitary rotation in Eq.~\ref{EQ:Unitary_p_to_q_rotation}), we can apply the Fourier Shift Theorem
\begin{equation} \label{eq:FShift}
    \mathcal{F} \{ g(x - x_o) \} = \frac{1}{2 \pi} \int dx \, g(x - x_o) e^{i x K} =  e^{i x_o K} \mathcal{F} \{ g(x) \},
\end{equation}
where $x_o$ is purely real and $K$ is the Fourier conjugate of $x$. By applying the results of Eq.~\ref{eq:FShift} to Eq.~\ref{Eq:V_RAD_shift}, we get
\begin{align} \label{Eq:V_RAD_phase}
    \hat{V}_\mathrm{RAD} =&~ \int \{ d {\bf K}_i \}\{ d {\bf K}_i' \}  ~ \, \exp \big( {-i \sum_j {\bf K}_j' \cdot \sum_\alpha {\boldsymbol{\xi}}_{ j,\alpha} \hat{q}_\alpha} \big) \\
    &~ \mathcal{V}(\{ {\bf K}_i' \}) \bigotimes_i  \ketbra{{\bf K}_i}{{\bf K}_i +{\bf K}_i'} \nonumber,
\end{align}
where $\mathcal{V}(K) = \mathcal{F} \{ {V} (x)\}$, and we introduced ${\bf K}_i' = {\bf K}_i'' -{\bf K}_i$. The origin of the name of this new Hamiltonian, Reciprocal Asymptotically Decoupled (RAD), is now apparent, since in the Fourier domain the light-matter interaction is mediated by a simple phase term $e^{-i \sum_{j,\alpha} {\bf K}_j' \cdot {\boldsymbol{\xi}_{j,\alpha}} \hat{q}_\alpha}$ in Eq.~\ref{Eq:V_RAD_shift} (as well as inside the kinetic energy term $\hat{T}_\mathrm{RAD}$). Using Eq.~\ref{Eqn:T_RAD} and Eq.~\ref{Eq:V_RAD_shift}, the total QED Hamiltonian can then be expressed for $N$ particles and $M$ modes as follows
\begin{widetext}
\begin{align} \label{eq:RAD_Ham}
\hat{H}_\mathrm{RAD}^{[N][M]} &=~ \sum_\alpha^M \hbar \Omega_\alpha (\hat{b}_\alpha^\dagger \hat{b}_\alpha + \frac{1}{2}) +\hat{T}_\mathrm{RAD} + \hat{V}_\mathrm{RAD} \\
\hat{T}_\mathrm{RAD}~&=~ \int \{ d {\bf K}_i \} \,\, \sum_j^N \bigg[ \frac{|\hbar {\bf K}_j|^2}{2 m_j} - \sum_{l,\alpha}^{N,M} \frac{\hbar^2 \Omega_\alpha^2}{2} ({\boldsymbol{\xi}}_{ j,\alpha} \cdot {\bf K}_j) ({\boldsymbol{\xi}}_{ l,\alpha} \cdot {\bf K}_l)  \bigg]
    \bigotimes_i^N \ketbra{{\bf K}_i}{{\bf K}_i}\nonumber \\
 \hat{V}_\mathrm{RAD} &= \int \{ d {\bf K}_i \}\{ d {\bf K}_i' \}  ~ \, \exp \bigg( -{i \sum_j^N {\bf K}_j' \cdot \sum_\alpha^M {\boldsymbol{\xi}}_{ j,\alpha} \hat{q}_\alpha} \bigg)\times \mathcal{V}(\{ {\bf K}_i' \}) \bigotimes_i^N  \ketbra{{\bf K}_i}{{\bf K}_i +{\bf K}_i'}.\nonumber 
\end{align}
\end{widetext}
%\begin{align} \label{eq:RAD_Ham}
% \hat{H}_\mathrm{RAD}^{[N][M]} =&~ \sum_\alpha^M \hbar \Omega_\alpha (\hat{b}_\alpha^\dagger \hat{b}_\alpha + \frac{1}{2}) +\hat{T}_\mathrm{RAD} + \hat{V}_\mathrm{RAD}  \\
% =&~ \sum_\alpha^M \hbar \Omega_\alpha (\hat{b}_\alpha^\dagger \hat{b}_\alpha + \frac{1}{2}) +\int \{ d {\bf K}_i \} \,\, \sum_j^N \bigg[ \frac{|\hbar {\bf K}_j|^2}{2 m_j} \nonumber\\
% &- \sum_{l,\alpha}^{N,M} \frac{\hbar^2 \Omega_\alpha^2}{2} ({\boldsymbol{\xi}}_{ j,\alpha} \cdot {\bf K}_j) ({\boldsymbol{\xi}}_{ l,\alpha} \cdot {\bf K}_l)  \bigg]
%     \bigotimes_i^N \ketbra{{\bf K}_i}{{\bf K}_i} \nonumber \\
% &+ \int \{ d {\bf K}_i \}\{ d {\bf K}_i' \}  ~ \, \exp \bigg( -{i \sum_j^N {\bf K}_j' \cdot \sum_\alpha^M {\boldsymbol{\xi}}_{ j,\alpha} \hat{q}_\alpha} \bigg) \nonumber \\
% &~~~ \times \mathcal{V}(\{ {\bf K}_i' \}) \bigotimes_i^N  \ketbra{{\bf K}_i}{{\bf K}_i +{\bf K}_i'}. \nonumber 
%\end{align}
The above QED Hamiltonian expression is the first {\it key result} of this paper. 
% For a model periodic potential $V(x)=v_{0}\cdot\cos(2\pi x/d)$, where $d$ is the lattice constant and $v_{0}$ is the potential depth, the corresponding $\mathcal{V}(K)=(v_0/2) [\delta(K-2\pi/d)+\delta(K+2\pi/d)]$ one can obtain the analytic expression of $\hat{H}_\mathrm{RAD}$ XXXXX 
The QED Hamiltonian $\hat{H}_\mathrm{RAD}$ in Eq.~\ref{eq:RAD_Ham} is general for any potential ${V}(x)$ (or its Fourier transform $\mathcal{V}(K)$, including {\it non-periodic} systems such as the potentials shown in Fig.~\ref{FIG:MAIN_POTS_WFNS_DIPS}. Compared to previous QED Hamiltonians (such as $\hat{H}_\mathrm{p \cdot A}$ in Eq.~\ref{eqn:Hc}, $\hat{H}_\mathrm{AD}$ in Eq.~\ref{eqn:AD_ham}, or the commonly used Pauli-Fierz Hamiltonian\cite{Mandal2023CR}), this new form has several advantages: (I) For periodic systems that are highly non-localized in space, their $\mathcal{V}(K)$ will be highly localized. (II) As shown in Fig.~\ref{FIG:MAIN_Xi_gc__loglog_Demler}a, the effective coupling parameter, $\xi_{j,\alpha}$, has a global maximum at a finite value of the actual coupling parameter $A_{0}$ (see Eq.~\ref{eqn:demlerHc}). This means that as long as the results converge for the highest value of $\xi_g$, they will converge for any arbitrary coupling strength (above or below) when using the RAD form of the QED Hamiltonian. (III) Compared to the original AD form of the QED Hamiltonian, $\hat{H}_\mathrm{RAD}$ in Eq.~\ref{eq:RAD_Ham} is guaranteed to work for any potential since the matrix elements only depend on the Fourier transform of the potential (\textit{i.e.}, no shifts by imaginary operators). 

It is insightful to check Eq.~\ref{eq:RAD_Ham} for different limits. If we consider a single particle ($N \xrightarrow{} 1$) in a potential interacting with many modes, the coupling parameter $\gamma_\alpha \to z |{\bf A}_\alpha| \sqrt{\omega_\alpha/m\hbar}$, and Eq.~\ref{eq:RAD_Ham} then simplifies to
\begin{align} \label{eq:h_rad-1par}
    \hat{H}_\mathrm{RAD}^{[1][M]} =&~ \sum_\alpha^M \hbar \Omega_\alpha (\hat{b}_\alpha^\dagger \hat{b}_\alpha + \frac{1}{2})  +\int d {\bf K} \,\, \frac{|\hbar {\bf K}|^2}{2 m_\mathrm{eff}}
    \ketbra{{\bf K}}{{\bf K}}  \\
    &+ \int d {\bf K} \,\, d {\bf K}'  ~ \, e^{ -{i {\bf K}' \cdot \sum_\alpha^M {\boldsymbol{\xi}}_{\alpha} \hat{q}_\alpha} } \mathcal{V}({\bf K}') \ketbra{{\bf K}}{{\bf K} +{\bf K}'}, \nonumber 
\end{align}
where all $\{i,j,l\}$ subscripts are removed, since there is only one particle, and $\hat{T}_\mathrm{RAD}$ is rewritten using the effective mass parameter defined by 
\begin{equation}\label{eq:m_eff}
    \frac{1}{m_\mathrm{eff}} =  \frac{1}{m} + \sum_\alpha \Omega_\alpha^2 \xi_\alpha \cos \phi_\alpha,
\end{equation}
where $\cos \phi_\alpha = {\bf A}_\alpha \cdot {\bf K} / |{\bf A}_\alpha || {\bf K}|$.

In the limit of a single mode and molecule ($\{N,M\} \xrightarrow{} 1$), this Hamiltonian further simplifies to
\begin{align} \label{eq:h_rad-1par-1mode}
    \hat{H}_\mathrm{RAD}^{[1][1]} =&~ \hbar \Omega (\hat{b}^\dagger \hat{b} + \frac{1}{2}) + \int d {\bf K} \,\,  \frac{|\hbar {\bf K}|^2}{2 m_\mathrm{eff}} \ketbra{{\bf K}}{{\bf K}}\\
    &+ \int d {\bf K} \,\, d {\bf K}'  ~ \, e^{ -{i {\bf K}' \cdot {\boldsymbol{\xi}_\mathrm{c}} \hat{q}_\mathrm{c}} } \mathcal{V}({\bf K}') \ketbra{{\bf K}}{{\bf K} +{\bf K}'}, \nonumber 
\end{align}
where the $\alpha$ subscripts are replaced by a $\mathrm{c}$ subscript indicating the single cavity mode. $\hat{q}_c$ is the photonic coordinate for the single mode, and $m_\mathrm{eff}$ is the effective mass of the dressed particle defined in Eq.~\ref{eq:m_eff} in the limit of a single mode. 

Due to the numerical cost and computational complexity involved for many-particle systems, all numerical results are presented for only the single-particle limit, as represented in Eqs.~\ref{eq:h_rad-1par} and \ref{eq:h_rad-1par-1mode}.

In order to expand on the RAD Hamiltonian's utility, we test its convergence on the two double-well models used in Ref.~\cite{Ashida2021PRL}, each in the single-particle and single-mode limit. Figs.~\ref{FIG:MAIN_POTS_WFNS_DIPS}a-b show the potential energies of a steep potential (panel a) and a shallow potential (panel b). The two models are defined as,
\begin{equation}
    V(x) = -\alpha x^2 + \beta x^4, \ \ \ \alpha, \beta > 0.
\end{equation}
The shallow potential ($\alpha = 3$, $\beta = 3.85$) is a good starting model for RAD, as the potential can be regarded as slowly varying; however, the steep potential ($\alpha = 50$, $\beta = 95$) is a harder test, since the potential will increase very quickly, effectively requiring more basis states to converge the result.

We will make a direct comparison to the well-known Pauli-Fierz (PF) Hamiltonian (see Appendix~\ref{app:PF} for the full derivation), which is a popular form of the QED Hamiltonian for computing polaritonic properties~\cite{Flick2017PNAS,Schaefer2018PRA,Rokaj2018JPBAMOP,Taylor2020PRL}. The PF Hamiltonian is expressed as follows
\begin{equation}\label{eq:PFham}
    \hat{H}_\mathrm{PF} = \hat{H}_M + \hbar \omega_c (\hat{d}^\dag \hat{d}+\frac{1}{2})+ \omega_c \hat{\mu}\cdot{\bf{A_0}} ( \hat{d}^\dag + \hat{d} ) + \frac{\omega_c}{\hbar} ({\bf{A_0}} \cdot \hat{\mu})^2,
\end{equation}
where $\hat{\mu}$ is the dipole operator of the matter and $\hat{d}$ is the photonic annihilation operator for the PF Hamiltonian in the dipole gauge, notably different from that of the annihilation operator of the Coulomb-gauge Hamiltonian due to a unitary transformation. Also note that ${\bf{A_0}} = \gamma_\mathrm{c} \sqrt{m\hbar/z^2 \omega_c} \hat{\bf{e}}$ is the vector potential vector for the single mode.

We solve the PF Hamiltonian by representing it in the eigenbases of $\hat{H}_\mathrm{M}$ and $\hbar \omega_c (\hat{d}^\dag \hat{d}+\frac{1}{2})$ (see Eq.~\ref{eq:PFham}) in a gauge-invariant truncated Hilbert space~\cite{Taylor2022OL,Taylor2020PRL,Bernardis2018PRAa} followed by a single-step numerical diagonalization. The matter eigenstates $|\psi_{\alpha}\rangle$ are obtained by diagonalizing the matter Hamiltonian directly using the discrete variable representation (DVR) \cite{colbert_novel_1992} (see Appendix~\ref{app:DVR}). The matter Hamiltonian $\hat{H}_\mathrm{M}$ is diagonalized with 2048 matter \textit{grid points} to provide the converged matter states and dipoles, which are then used as the input to diagonalize $\hat{H}_\mathrm{PF}$ (Eq.~\ref{eq:PFham}). For the RAD Hamiltonian (Eq.~\ref{eq:h_rad-1par-1mode}), $V(x)$ is represented in a much smaller 100 matter grid point basis, and we then perform an asymmetrically normalized FFT (as in Eq.~\ref{eq:FT}) to represent it in the eigenbasis of $\hat{p}$. By expressing the photonic DOF in the eigenbasis of $\hbar \Omega (\hat{b}^\dagger \hat{b} + \frac{1}{2})$ and the matter DOF in the eigenbasis of $\hat{p}$, we can directly diagonalize the total Hamiltonian. Due to this, there is no need to precompute the matter eigenstates $|\psi_{\alpha}\rangle$ or the dipole matrix elements for RAD.

Figs.~\ref{FIG:MAIN_Xi_gc__loglog_Demler}b,c showcase the convergence of the RAD and PF Hamiltonians, with the eigenspectrum of the two double-well potentials as functions of the normalized coupling strength $\gamma_\mathrm{c}/\omega_c$. Both panels are plotted on a log-log scale, where the PF results are shown in open circles, and the RAD results are shown in thick solid lines. The convergence of the two Hamiltonians is the focus of the discussion, since they are formally equivalent and related to each other through a unitary transformation. Since the RAD Hamiltonian is expected to converge faster than the PF Hamiltonian at large coupling strengths, the PF will require much more matter and Fock states to converge the results. Additionally, the steep potential (Fig.~\ref{FIG:MAIN_POTS_WFNS_DIPS}b) was expected to be more challenging for the RAD Hamiltonian, since we solve this Hamiltonian in reciprocal space. Nonetheless, the convergence for the RAD Hamiltonian was achieved using 100 matter \textit{grid points} and 20 RAD Fock states (eigenstates of $\hbar \Omega (\hat{b}^\dagger \hat{b} + \frac{1}{2})$ in Eq.~\ref{eq:h_rad-1par-1mode}), while the PF Hamiltonian required 50 matter states and 200 PF Fock states (eigenstates of $\hbar \omega_c (\hat{d}^\dag \hat{d}+\frac{1}{2})$ in Eq.~\ref{eq:PFham}). To be clear, the meaning of a RAD Fock state and a PF Fock state is not the same due to the different gauges used for the photonic operators $\hat{b}$ and $\hat{d}$, respectively, and can be converted to one another (and to the p$\cdot$A) via a unitary gauge transformation. However, our discussion is primarily concerned with the convergence with the number of Fock states regardless of the gauge (or representation), so we will not explicitly distinguish between RAD and PF Fock states and simply refer to both as Fock states. We note that the results from diagonalizing the PF Hamiltonian do not yet match those of the RAD, but increasing the basis further is not computationally feasible for realistic calculations. However, it is already clear that the factor of 5 reduction in the size of the converged basis (or $1/25$ the number of matrix elements) is enough proof of the feasibility of the RAD Hamiltonian.

% We want to further discuss how the matter grid points and converged matter states do not contain the same meaning. The RAD Hamiltonian requires solving the entire Hamiltonian once with a certain number of $K$-space grid points and RAD Fock states. The PF Hamiltonian, on the other hand, often needs the bare-matter energies and electronic dipole matrix elements between matter eigenstates. As such, for the PF Hamiltonian, one may solve the matter part independently of the photonic DOFs, and then diagonalize the full Hamiltonian using these matter eigenstates. This allows one to fully converge the matter energies and dipoles up to the required degree. For RAD, only the raw electronic potential is used for input into the RAD Hamiltonian, favoring a self-consistent approach to the solution of the, in principle, many-body Hamiltonian, as is often done for the PF Hamiltonian for many-electron systems.\cite{Flick2017PNAS,Schaefer2018PRA,Rokaj2018JPBAMOP,haugland_QEDIntermol_JCP2021,haugland_QEDCC_PRX2020,riso_QEDHF_NatCommun2022,riso_QEDIONIZATION_JCP2022,vu_enhanced_JPCA2022,deprince_QEDIONIZATION_JCP2021,liebenthal_QEDEOMCCSD_JCP2022,mctague_nhQEDCIS_JCP2022}

Fig.~\ref{FIG:MAIN_Xi_gc__loglog_Demler}c shows results for the shallow potential (Fig.~\ref{FIG:MAIN_POTS_WFNS_DIPS}b) using the same convergence parameters as in Fig.~\ref{FIG:MAIN_Xi_gc__loglog_Demler}b. For RAD, however, only five Fock states were required to converge the results instead of 20, implying that the shallow (slowly varying) potential is much easier for the RAD Hamiltonian to converge. However, the PF Hamiltonian is not converged with 50 matter states and 200 Fock states, which indicates that the PF Hamiltonian struggles to converge at these large coupling strengths for both models in panels b and c. Appendix~\ref{app:fock_states} shows additional numerical results that provide insight into the contribution of higher-energy Fock states to polariton states in both the RAD and the PF representations.

\section{Extension to Periodic Systems and Polariton Band Structure} \label{sec:period}
The primary objective of this work is to provide a rigorous and efficient Hamiltonian tailored for periodic systems. Without losing generality, let us consider a 3-D periodic potential function with three direct lattice basis vectors, $\{ {\bf a}_1, {\bf a}_2, {\bf a}_3 \}$, and a spatially localized function $v(\{ {\bf x}_i \})$ for one unit cell, such that the periodic potential is expressed as
\begin{equation}\label{V-pot-conv-3D}
    V(\{ {\bf x}_i \}) = v(\{ {\bf x}_i \}) \circledast \prod_{i,\nu} \Sh \left( \frac{{\bf x}_i \cdot {\bf a}_\nu}{|{\bf a}_\nu|^2} \right),
\end{equation}
where $\circledast$ denotes a convolution between two functions $f(x)\circledast g(x)=\int_{-\infty}^{\infty} f(x')g(x-x')dx'$, $\Sh(x)$ is the Dirac comb function, and $\{ \nu \} \in \{ 1,2,3\}$ iterates over the direct lattice basis vectors. 

Using the Fourier transform definition in Eq.~\ref{eq:FT}, as well as the Fourier Convolution Theorem
\begin{equation}
    \mathcal{F}\{ g(x) \circledast h(x) \} = 2 \pi \mathcal{F}\{ g(x) \} \mathcal{F}\{ h(x) \},
\end{equation}
we can obtain the Fourier Transformed potential $\mathcal{V}(\{ {\bf K}_i \})$ as
\begin{equation} \label{eq:period_potential-3D}
    \mathcal{V}(\{ {\bf K}_i \}) =  {v}(\{ {\bf K}_i \}) \cdot \prod_{i,\nu} \Sh \left( \frac{{\bf K}_i \cdot {\bf b}_\nu} {|{\bf b}_\nu|^2} \right),
\end{equation}
where $\{ {\bf b}_\nu \}$ are the reciprocal lattice basis vectors. Eq.~\ref{eq:period_potential-3D} can also be thought of as an implicit restatement of Bloch's theorem (see Appendix \ref{app:bloch}). One convenient way to define $v(K)$ is
\begin{equation} \label{eq:v_k_expr-3D}
    v(K) = \frac{1}{8 \pi^3 {V}_\mathrm{cell}} \int_{{V}_\mathrm{cell}} \{d{\bf x}_i\} \,\, V(\{{\bf x}_i\}) e^{i \sum_i {\bf K}_i \cdot {\bf x}_i},
\end{equation}
where the above expression is the Fourier Transform of $V(\{{\bf x}_i\})$ over a single unit cell, normalized by the volume of the unit cell, $V_\mathrm{cell}$. However, in principle, $v(\{{\bf K}_i\})$ is not unique for a given $V(\{{\bf x}_i\})$, but the representation of the potential in Eq.~\ref{eq:v_k_expr-3D} is often used and is easily accessible in electronic structure calculations aimed at simulating periodic systems. The inclusion of pseudopotentials, coupled with the projector augmented wave method, for the core electrons in such packages, adds an additional complication but in principle can still be cast in this or a similar form.\cite{Bloechl1994PRB, Kresse1999PRB}

By inserting the potential from Eq.~\ref{eq:period_potential-3D} into Eq.~\ref{Eq:V_RAD_phase}, $\hat{V}_\mathrm{RAD}$ becomes
\begin{align} \label{eq:period_v_rad}
    \hat{V}_\mathrm{RAD} =&~ \sum_{\{ {\boldsymbol{\kappa}}_i' \}} \int \{ d {\bf K}_i \} ~ \, \exp \big( -{i \sum_j {\boldsymbol{\kappa}}_j' \cdot \sum_\alpha {\boldsymbol{\xi}}_{ j,\alpha} \hat{q}_\alpha} \big) \\
    &~ v(\{ {\boldsymbol{\kappa}}_i' \}) \bigotimes_i  \ketbra{{\bf K}_i}{{\bf K}_i + {\boldsymbol{\kappa}}_i'} \nonumber,
\end{align}
where $\{ {\boldsymbol{\kappa}}_i' \} \in \{ \sum_\nu n_\nu {\bf b}_\nu \}, \forall~ n_\nu \in \mathbb{Z}$ is the set of reciprocal lattice vectors, formed due to the sifting property of the Dirac comb function.

For periodic systems, dispersion plots are more insightful for characterizing the system compared to the eigenspectrum of the full Hamiltonian. To calculate the polaritonic dispersion plot of a system described by $\hat{H}_\mathrm{RAD}$, we consider the Hamiltonian for each $k$-point in the first Brillouin zone as $\hat{H}_\mathrm{RAD}({\bf k})$,
% \begin{equation}
% \hat{H}_\mathrm{RAD}({\bf k}) = \hat{P}_{\bf k} \hat{H}_\mathrm{RAD}  \hat{P}_{\bf k}
% \end{equation}
where it is confined by a projection operator $\hat{\mathcal{P}}_k$, a global operator that projects all degrees of freedom (matter and photonic) to a given $k$-point. We define this projection operator $\hat{\mathcal{P}}_{\bf k}$, where ${\bf k}$ is confined to the matter's first Brillouin zone, as
\begin{align} \label{eq:p_k}
    \hat{\mathcal P}_{\bf k} &= \hat{\mathcal P}_{\bf k}^\mathrm{el} \otimes \hat{\mathcal P}_{\bf k_\beta}^\mathrm{ph} \\
    &= \nonumber \bigotimes_i \left( \sum_{\{ \boldsymbol{\kappa}_i \} } \ketbra{{\bf k} + \boldsymbol{\kappa}_i}{{\bf k} + \boldsymbol{\kappa}_i} \right) \otimes \left( \sum_{n_{\beta}} \ketbra{n_{\beta}}{n_{\beta}} \right),
\end{align}
where for simplicity, we are only considering the transverse electric (TE) polarization, such that $\beta = \{ {\bf k}_\beta, \mathrm{TE}\}$ and $\ket{n_\beta}$ is a Fock state of excitation $n$ with the wavevector, ${\bf k}_\beta = \bf k$ with a TE polarization direction. Note that this choice of ${\bf k}_\beta = \bf k$ is just a cross-section of, in principle, a two-dimensional dispersion relation, but it still allows us to extract physical insight into how coupling to cavity changes the properties of the system. Additionally, this projection operator confines each ${\bf K}'_i$ to the values of ${\bf k}_i+\boldsymbol{\kappa}_i$, and $\hat{\mathcal{P}}^2_k=\hat{\mathcal{P}}_k$. In doing this, we change ${\bf K}'_i \to {\bf k}_i+\boldsymbol{\kappa}_i$, where $\boldsymbol{\kappa}_i \in \{ \sum_\nu n_\nu {\bf b}_\nu \}, \forall~n_\nu \in \mathbb{Z} $ is also a reciprocal lattice vector. 

As discussed in Ref.~\citenum{Taylor2022OL}, one must be careful of gauge ambiguities when applying a projection of photonic modes. To generate $\hat{H}_\mathrm{RAD}({\bf k})$, we first truncate Eq.~\ref{eqn:Hc} using $\hat{\mathcal{P}}_{\bf k}^\mathrm{ph}$. For a single $\bf k$ term, it should be noted that including both polarizations is a simple extension, since $\hat{\bf e}_{{\bf k}, \lambda} \cdot \hat{\bf e}_{{\bf k}, \lambda'} = \delta_{\lambda,\lambda'}$. In this special case of a single $\bf k_\beta$, the normal mode transformation simplifies to a Bogoliubov transformation (see Appendix~\ref{app:bogoliubov}). The rest of the RAD derivation then follows accordingly from Eq.~\ref{eq:shift_pA} onward. In this case, $\hat{H}_\mathrm{RAD}({\bf k})$ is generated by projecting $\hat{H}_\mathrm{RAD}$, which is properly confined in $\hat{\mathcal{P}}_{\bf k}^\mathrm{el} \otimes \hat{\mathcal{P}}_{{\bf k}_\beta}^\mathrm{ph}$. Applying this simple projection by $\hat{\mathcal{P}}_{\bf k}^\mathrm{el}$ does not cause any gauge ambiguities, since it is done in the eigenbasis of $\{ \hat{\bf p}_j \}$, essentially writing the Hamiltonian in momentum space using a grid-basis. 

Using the form of $\hat{V}_\mathrm{RAD}$ from Eq.~\ref{eq:period_v_rad}, we can express $\hat{H}_\mathrm{RAD} ({\bf k})$ (the ${\bf k}$-resolved RAD) as,
\begin{widetext}
\begin{align}\label{eq:h_rad_k_gen}
    \hat{H}_{\mathrm{RAD}} ({\bf k} = {\bf k}_\beta) =&\sum_{\{ \boldsymbol{\kappa}_i \}} \sum_j \bigg[ \frac{|\hbar ({\bf k} + \boldsymbol{\kappa}_j)|^2}{2 m_j} - \sum_{l} \frac{\hbar^2 \Omega_{\beta}^2}{2} ({\boldsymbol{\xi}}_{ j,\beta} \cdot ({\bf k} + \boldsymbol{\kappa}_j)) ({\boldsymbol{\xi}}_{ l,\beta} \cdot ({\bf k} + \boldsymbol{\kappa}_l))  \bigg]\bigotimes_i^N \ketbra{{\bf k} + \boldsymbol{\kappa}_i}{{\bf k} + \boldsymbol{\kappa}_i} \\
    &+ \sum_{\{\boldsymbol{\kappa}_i,\boldsymbol{\kappa}_i'\}} \exp \bigg( -{i \sum_j \boldsymbol{\kappa}_j' \cdot \sum_\lambda {\boldsymbol{\xi}}_{ j,\beta} \hat{q}_{\beta} } \bigg)  v(\{ \boldsymbol{\kappa}_i' \}) \bigotimes_i  \ketbra{ {\bf k} + \boldsymbol{\kappa}_i }{{\bf k} + \boldsymbol{\kappa}_i +\boldsymbol{\kappa}_i'} + \hbar \Omega_{\beta} (\hat{b}_{\beta}^\dagger \hat{b}_{\beta} + \frac{1}{2}), \nonumber  
\end{align}
\end{widetext}
where all the integrals have now been replaced by discrete sums, creating a drastic decrease in the size of the Hilbert space (See Eq.~\ref{eq:p_j_i} for a simplified example of the indexing in this many-particle and many mode Hilbert space). By solving the eigenspectrum of Eq.~\ref{eq:h_rad_k_gen} for each ${\bf k}$-point, we can form the dispersion plot of the system. The above expression of $\hat{H}_\mathrm{RAD}({\bf k})$ is the second {\it key result} of this paper. 

% The form in Eq.~\ref{eq:h_rad_k_gen} is general for an arbitrary potential, number of particles, and number of modes to the same level of theory as the minimal coupling Hamiltonian. Once again, for the purposes of this paper, we will test this Hamiltonian on single-particle, one-dimensional models with a single $\lambda$. In this limit, $\hat{H}_{\mathrm{RAD}}^{\bf k}$ can be expressed as
% \begin{align}
%     \hat{H}_{\mathrm{RAD}}^{\bf k} =&~ \hbar \Omega_{ k} (\hat{b}_{{ k}}^\dagger \hat{b}_{{ k}} + \frac{1}{2}) + \sum_{{\kappa}} \frac{|\hbar ({ k} + {\kappa})|^2}{2 m_\mathrm{eff}} \ketbra{{ k}+ {\kappa}}{{ k} + {\kappa}} \nonumber \\
%     &+ \sum_{{\kappa} , {\kappa}'} e^{ {i {\kappa}' \cdot {{\xi}}_k \hat{q}_{k}} } v({\kappa}') \ketbra{{ k}+ {\kappa}}{{ k} + {\kappa} + {\kappa}'}.  
% \end{align}
% With this concise representation of the RAD Hamiltonian in this limit, we are now prepared to perform numerical simulations on model potentials.

% While we demonstrated in the previous section that the RAD Hamiltonian can accurately simulate spatially localized Hamiltonians for arbitrarily coupling strengths, the design intent for this method is for periodic systems. We will now demonstrate the performance of the RAD Hamiltonian with a novel periodic modified Coulomb model potential. This model is beyond the capabilities of the AD frame, requiring the RAD Hamiltonian representation. 

\begin{figure}
 \centering
 %%-------start of first figure----------
    \begin{minipage}[h]{\linewidth}
     \centering
     \includegraphics[width=1.0\linewidth]{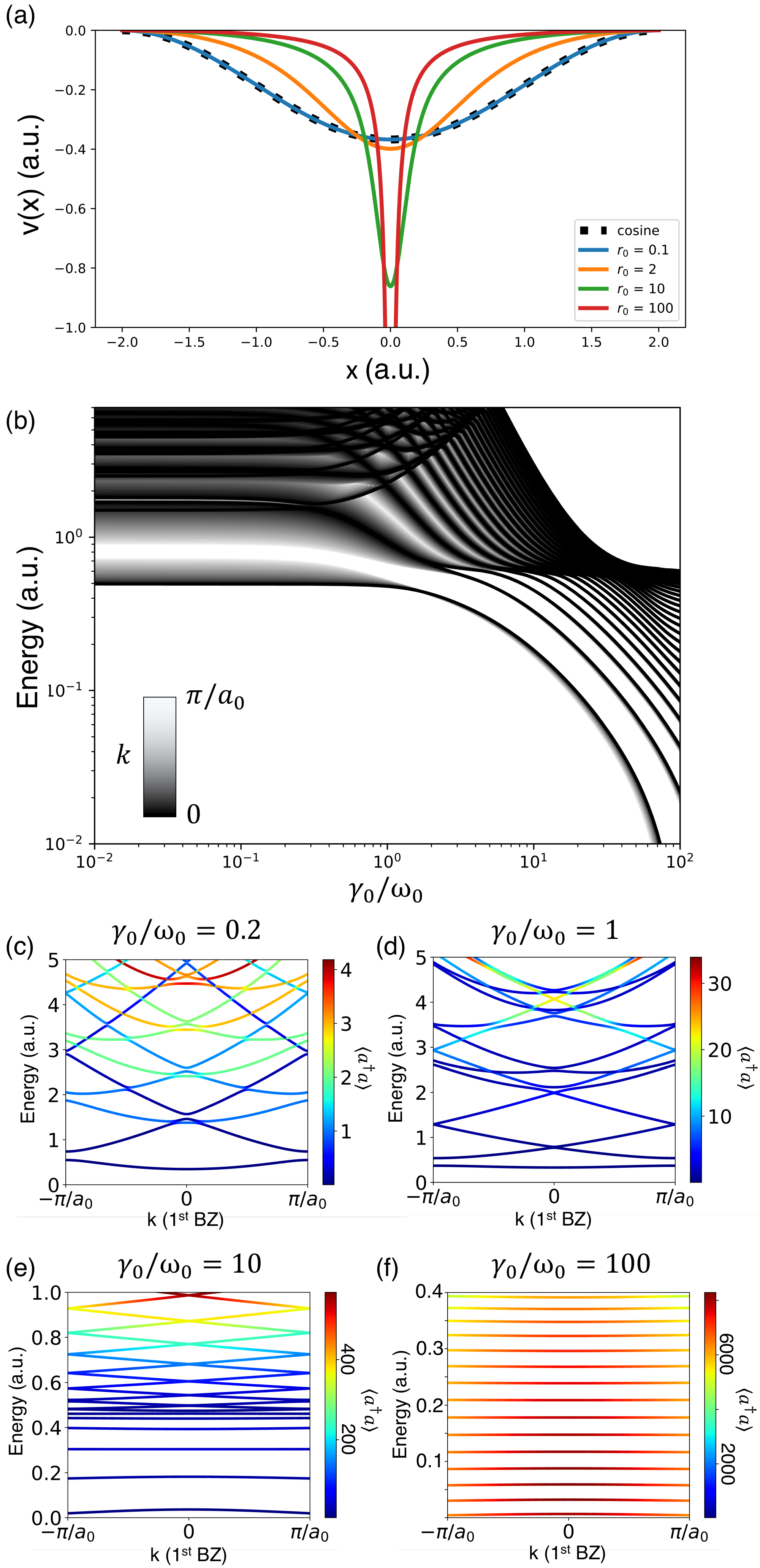}
    \end{minipage}%
   \caption{ Single electron in a periodic modified Coulomb potential coupled to many longitudinal cavity modes. (a) Single unit cell of the periodic modified Coulomb potential for many $r_0$ values plotted upon a cosine potential, where the $Z$ for the modified Coulomb potential is adjusted such that the modified Coulomb potential's first order Fourier expansion matches the cosine potential (b) First 40 bands in the energy eigenspectrum of this model as a function of normalized coupling strength $\gamma_0/\omega_0$ color-coded by $k$-point (where $\gamma_0$ and $\omega_0$ are the coupling and cavity frequency at $k=0$). (c)-(f) shows the dispersion plots for normalized coupling strengths of $\gamma_0/\omega_0 = 0.2, 1, 10, 100$, respectively, where the color shows the expectation value of the photon number in the Coulomb Gauge, $\langle \hat{a}^\dagger \hat{a} \rangle$. }
   \label{fig:erf_pot}
\end{figure}

\section{Numerical Results on Model 1D Periodic System}\label{sec:1Derf}
While we demonstrated in Sec.~\ref{sec:RAD} that the RAD Hamiltonian can accurately simulate spatially localized Hamiltonians for arbitrarily coupling strengths, the intent of this method is for periodic systems. We will now demonstrate the performance of the RAD Hamiltonian with a model periodic modified Coulomb potential. This model is beyond the capabilities of the AD Hamiltonian~\cite{Ashida2021PRL}, explicitly requiring the use of the RAD Hamiltonian.

As we will simulate this RAD framework with many photonic modes, it is important to define the dispersion relation for the cavity. For the sake of simplicity, we assume that the cavity dispersion takes the form of the typical Fabry-P\'erot cavity, and the 1D matter chain is parallel to the cavity mirrors. In such a case, the cavity dispersion takes the form of
\begin{equation} \label{eq:disp_rel}
    \omega_{k,\lambda} = \sqrt{\omega_\mathrm{c}^2 + c^2 k^2},
\end{equation}
where $c$ is the speed of light inside the cavity.

Although the form in Eq.~\ref{eq:h_rad_k_gen} is general for an arbitrary potential, number of particles, and number of modes to the same level of theory as the minimal coupling Hamiltonian, for the purposes of this paper, we will test this Hamiltonian on a single-particle, one-dimensional model with a single $\lambda$. In this limit, $\hat{H}_{\mathrm{RAD}}(k)$ can be expressed as
\begin{align}
    \hat{H}_{\mathrm{RAD}} ({k}) =&~ \hbar \Omega_{ k} (\hat{b}_{{ k}}^\dagger \hat{b}_{{ k}} + \frac{1}{2}) + \sum_{{\kappa}} \frac{|\hbar ({ k} + {\kappa})|^2}{2 m_\mathrm{eff}} \ketbra{{ k}+ {\kappa}}{{ k} + {\kappa}} \nonumber \\
    &+ \sum_{{\kappa} , {\kappa}'} e^{ -{i {\kappa}' \cdot {{\xi}}_k \hat{q}_{k}} } v({\kappa}') \ketbra{{ k}+ {\kappa}}{{ k} + {\kappa} + {\kappa}'},  
\end{align}
where $\{\kappa, \kappa'\} \in \{ 2 \pi n / a_0 \}, \forall~ n \in \mathbb{Z}$.
With this concise representation of the RAD Hamiltonian in this limit, we are now prepared to perform numerical simulations on model potentials.

We define a periodic array of modified Coulomb potentials.  We build this model based on the famous Shin-Metiu molecular model~\cite{Shin1995JCP}. We approximate the potential of each ion as a modified Coulomb potential of $V(x) = - Z e^2 \mathrm{erf}(r_0 x)/x$, where $Z$ is the effective charge of the ion and $r_0$ is a parameter to determine the steepness of the potential. Then, we place these ions on an infinite 1D lattice, separated by the lattice constant, $a_0$. The localized potential then has the form,
\begin{align}
    v({x}) &=  - \frac{ Z e}{a_0} \frac{\mathrm{erf} \left( {r_0 x}\right)}{x}, \\
    v(K) &= -\frac{Z e^2}{2 \pi} \Gamma\bigg( 0 , \left[ \frac{K}{2 r_0} \right]^2 \bigg) 
\end{align}
where $\Gamma(0,x)$ is the $0^\mathrm{th}$ order upper incomplete gamma function. The Dirac comb function turns one of the integrals in Eq.~\ref{eq:RAD_Ham} into a sum. The parameter, $r_0$, can drastically affect the shape of this periodic potential with interesting limits. In the limit of $r_0 \to 0$, $v(x) \propto \cos(a_0 x)$, and in the limit of $r_0 \to \infty$, the potential approaches a delta function. Fig.~\ref{fig:erf_pot}a shows a single period of $v(x)$ for various different values of $r_0$, where the value of $Z$ is varied such that the magnitude of the first coefficient in the Fourier expansion is the same for each potential plotted and matches the reference cosine curve plotted (black dotted line). 

The total Hamiltonian of this system then becomes,
\begin{align} 
    &\hat{H}_\mathrm{RAD} =  \, \sum_\alpha \hbar \Omega_\alpha \Big(\hat{b}^\dagger_\alpha \hat{b}_\alpha + \frac{1}{2} \Big) + \int dK \, \frac{\hbar K}{2 m_\mathrm{eff}} \ketbra{K}{K} \\
    &-\frac{Z e^2}{2 \pi} \sum_{\kappa' \neq 0} \int dK \, \, \ketbra{K}{K +\kappa'}  \, e^{-i \kappa' \sum_\alpha {{\xi}}_{\alpha} \hat{q}_\alpha} \, \, \Gamma\left( 0 , \left[ \frac{\kappa'}{2 r_0} \right]^2 \right) \nonumber,
\end{align}
where the sum over $\kappa'$ in theory goes to $\pm \infty$, but in practice the upper incomplete gamma function decays so fast that only a few values of $\kappa'$ are significant for typical values of $a_0$. Note that the sum over $\kappa'$ does not include $\kappa' = 0$ since that term is in general a zero-point energy shift, and in this case, it is a singularity for the upper gamma function. The $k$-resolved form of this Hamiltonian can then be written as
\begin{align} \label{eq:H_k_RAD_gamma}
    &\hat{H}_\mathrm{RAD} (k) =  \, \hbar \Omega_k (\hat{b}^\dagger_k \hat{b}_k + \frac{1}{2})+ \sum_{\kappa} \frac{\hbar K}{2 m_\mathrm{eff}} \ketbra{ k +\kappa}{k +\kappa} \\
    &-\frac{Z e^2}{2 \pi}  \sum_{\kappa} \sum_{\kappa' \neq 0} \ketbra{k + \kappa}{k + \kappa +\kappa'}  \, e^{-i \kappa' {\xi_k} \hat{q}_k} \, \, \Gamma\left( 0 , \left[ \frac{\kappa'}{2 r_0} \right]^2 \right) \nonumber.
\end{align}
The Hamiltonian in Eq.~\ref{eq:H_k_RAD_gamma} is now in a form that can be easily diagonalized to find the eigenenergies of this system.

Figs.~\ref{fig:erf_pot}b-f presents the {\it polaritonic} dispersion obtained by solving the eigenvalues of Eq.~\ref{eq:H_k_RAD_gamma} for different $k$-points. Each point on these polaritonic dispersion plots is obtained by projecting both the matter and the photonic DOFs to the same $k$-point (cf. Eq.~\ref{eq:p_k}). Although this provides a concise description of the energetic structure of this hybrid system, it should be noted that, in general, this is only a slice through the two-dimensional dispersion surfaces (for matter $k$ and photonic $k_\beta$), to be discussed later in Fig.~\ref{fig:blwa_2d}. Regardless, this cross-section still allows us to extract valuable insights into how this 1D matter system is modified by its coupling to the cavity. All the results for panels Fig.~\ref{fig:erf_pot}b-f converged completely for 5 Fock states and 101 $\kappa$ values.

Fig.~\ref{fig:erf_pot}b shows how these polaritonic dispersions change with increasing coupling strength, where the value of $k$ is represented by the shade of the lines. Since, in principle, $\gamma_\beta$ and $\omega_\beta$ are $k$-dependent, we define the coupling strength of the system by $\gamma_0 / \omega_0$, where the 0 subscript refers to these values at the gamma point. This fixes the couplings and frequencies for all other modes based on the dispersion relation in Eq.~\ref{eq:disp_rel}. As the coupling strength goes into the deep strong coupling regime, the bands flatten (\textit{i.e.}, each band loses its dependence on $k$), and the total density of states becomes sharply peaked. This band-flattening can be more clearly seen in panels c-f (note the change in vertical scales between panels) where the dispersion is shown at various choices of coupling strength. Intuitively, this can be understood by considering how $\hat{T}_\mathrm{RAD}$ (Eq.~\ref{Eqn:T_RAD}) is affected by the coupling strength. As $\gamma_0 \to \infty$, the effective mass $m_\mathrm{eff} \to \infty$, making $\hat{T}_\mathrm{RAD} \to 0$. This makes the matter dispersion of the light-dressed particle flatten. Note that this is directly opposite to the free-electron model, which only contains the kinetic energy. Additionally, in this limit of $\hat{T}_\mathrm{RAD} \to 0$, the commutation between $\hat{H}_\mathrm{RAD}$ and the matter momentum, $\hat{p}$, approaches zero. This allows the polaritonic bands to become arbitrarily close together and eventually degenerate.

To take a closer look at how these polaritonic dispersion plots behave at different coupling strengths, Figs.~\ref{fig:erf_pot}c-f present the dispersion curves of the polaritonic band structure at various coupling strengths $\gamma_0 / \omega_0$, and each panel is a cross-section of the more general plot shown in Fig.~\ref{fig:erf_pot}b.  The colors in Figs.~\ref{fig:erf_pot}c-f now represent the expectation value of the physical photon number in the Coulomb gauge, $\langle \hat{a}^\dagger \hat{a} \rangle$ for each state. Recall that the results were converged using only five RAD Fock states and were then transformed by unitary rotation to the Coulomb gauge, which we take as the physical photon number. When $\gamma_0/\omega_0 = 0.2$ (Fig.~\ref{fig:erf_pot}c), the dispersion relation appears as one would expect in the strong coupling regime, where the matter bands are duplicated and shifted up in energy by $\omega_k$ for each added photon. When bands of different photon numbers cross, there is Rabi splitting. As the coupling increases to $\gamma_0/\omega_0 = 1$, the standard intuition from the strong coupling regime no longer applies. The band structure is almost completely different from the uncoupled case. Additionally, for bands in the same energy range as the lower coupling case (Fig.~\ref{fig:erf_pot}c), the average photon number is much higher, with some bands reaching an $\langle \hat{a}^\dagger \hat{a} \rangle > 30$. This effect is magnified as the coupling goes further into the deep strong coupling regime with the lowest-energy band for $\gamma_0/\omega_0 = 100$ having $\langle \hat{a}^\dagger \hat{a} \rangle > 8000$ for some $k$-points. We again emphasize that the $\langle \hat{a}^\dagger \hat{a} \rangle$ plotted is the Coulomb-gauge photon number, which is the quantity accessible to experiments and not directly related to the photon basis used for RAD. This is a key benefit of the RAD representation.
%We emphasize that the $\langle \hat{a}^\dagger \hat{a} \rangle$ plotted is the photon number in the Coulomb gauge. The results obtained using the RAD representation converge with only 5 Fock states. 

This 1D-modified Coulomb potential model demonstrates the strengths of the RAD Hamiltonian representation. By modeling this matter system in the strong, ultrastrong, and deep strong regimes, this RAD representation provides both a convenient intuitive understanding of seemingly unnatural results and a computationally efficient basis to easily converge numerical simulations.

\section{Beyond the Long Wavelength Approximation} \label{sec:BLWA}
In Section~\ref{sec:AD}, we began our discussion by assuming the long-wavelength approximation (LWA) in the Coulomb gauge Hamiltonian (Eq.~\ref{eqn:demlerHc}). This approximation assumes that the spatial variation of the vector potential field across the matter system is small enough to be considered negligible. For a simple Fabry-P\'erot cavity, this approximation is written as
\begin{equation} \label{eq:A_BLWA}
    \hat{\bf A}({\bf r}) = \sum_{\beta} {\bf A}_{{\beta}} ( \hat{a}_{\beta}^\dagger e^{- i {\bf k}_\beta \cdot {\bf r}} + \hat{a}_{\beta} e^{i {\bf k}_\beta \cdot {\bf r}}) \to \hat{\bf A} = \sum_{\beta} {\bf A}_{\beta} ( \hat{a}_{\beta}^\dagger + \hat{a}_{\beta} ),
\end{equation}
where $\beta$ is the superindex that indexes over all $\bf k_\beta$ and $\lambda$, and $\bf r$ is the spatial coordinate of the cavity.

In this section, we start to relax this approximation for a single particle coupled to many spatially varying modes in a Fabry-P\'erot (FP) type cavity with the dispersion relation
\begin{equation}
    \omega_\beta = \sqrt{\omega_\mathrm{c}^2 + c^2 |{\bf k}_\beta|^2},
\end{equation}
where we set ${\bf k}_\beta$ as the component of the photonic wavevector that is parallel to the cavity mirrors.
While experimentally, matter coupled to FP cavities has not reached the deep-strong coupling regime, the plane-wave basis of the EM field allows a convenient way to model the spatial variations of the vector potential and still allows us to glean valuable physical insights from our results. In principle, any arbitrary cavity can be represented on a plane-wave basis, but for simplicity, we consider an FP cavity in this work.

We begin by expressing the Coulomb gauge Hamiltonian for a single particle while using the exact form of $\hat{\bf A}$ from Eq.~\ref{eq:A_BLWA} as
\begin{align} \label{eq:h_pa_blwa}
    \hat{H}_\mathrm{p \cdot A} =&~ \hat{H}_\mathrm{M} +  \sum_\beta \hbar \omega_\beta (\hat{a}_\beta^\dagger \hat{a}_\beta + \frac{1}{2})  \\
    &- \sum_{ \beta} \frac{z_j \hat{\bf p} \cdot {\bf A}_\beta}{m} (\hat{a}^\dagger_\beta e^{- i {\bf k}_\beta \cdot \hat{\bf x}} + \hat{a}_\beta  e^{i {\bf k}_\beta \cdot \hat{\bf x}})  
    +  \frac{z^2 |\hat{\bf A} |^2 }{2 m}. \nonumber
\end{align}
Following the same procedure as before (Eq.~\ref{eq:HPA-gc} - Eq.~\ref{eq:qp_beta}), we now define the $\beta_\mathrm{th}$ mode's spatially varying photonic coordinate ($\hat{\Tilde{q}}_\beta$) and momentum ($\hat{\Tilde{p}}_\beta$) operators as
\begin{subequations}
\begin{align}
\hat{\Tilde{q}}_\beta (\hat{\bf x}) = \sqrt{\frac{\hbar}{2\omega_\beta}}(\hat{a}^\dagger_\beta e^{- i {\bf k}_\beta \cdot \hat{\bf x}} + \hat{a}_\beta  e^{i {\bf k}_\beta \cdot \hat{\bf x}})\\
\hat{\Tilde{p}}_\beta (\hat{\bf x}) = i\sqrt{\frac{\hbar\omega_\beta}{2}}(\hat{a}^\dagger_\beta e^{- i {\bf k}_\beta \cdot \hat{\bf x}} - \hat{a}_\beta  e^{i {\bf k}_\beta \cdot \hat{\bf x}}). 
\end{align}
\end{subequations}
These operators $\hat{\Tilde{q}}_\beta$ and $\hat{\Tilde{p}}_\beta$ maintain the same commutation relations as $\hat{q}_\beta$ and $\hat{p}_\beta$, and $\hat{H}_\mathrm{ph} = \sum_\beta \hat{p}_\beta^2 + \omega_\beta \hat{q}_\beta^2 = \sum_\beta \hat{\Tilde{p}}_\beta^2 + \omega_\beta \hat{\Tilde{q}}_\beta^2$. By moving the $\hat{\bf x}$ dependence within the definitions of $\hat{\Tilde{q}}_\beta$ and $\hat{\Tilde{p}}_\beta$, the normal mode analysis done in Section~\ref{sec:AD} and Appendix~\ref{app:normal_modes} is not affected by relaxing the LWA. As such, the expression for $\hat{T}_\mathrm{AD}$ in Eq.~\ref{eq:T_AD} is unaffected by the LWA. We can then write the Coulomb gauge Hamiltonian beyond the long-wavelength approximation after a normal mode transformation as
\begin{align}
    \hat{H}_\mathrm{p \cdot A} =&~ \hat{T}_\mathrm{AD} + \hat{V}({\bf x}) + \sum_\alpha \frac{1}{2} \Big( \hat{\Tilde{p}}_\alpha(\hat{\bf x})^2 + \Omega_\alpha^2 (\hat{\Tilde{q}}_\alpha(\hat{\bf x}) - \hat{\bf p} \cdot \boldsymbol{\xi}_\alpha )^2 \Big),
\end{align}
where the shift of $\hat{\Tilde{q}}_\alpha(\hat{\bf x})$ by $\hat{\bf p} \cdot \boldsymbol{\xi}_\alpha$ in the Coulomb gauge, is now explicitly written. Note that the corresponding LWA expression is in Eq.~\ref{eq:shift_pA}.

Further, the $\hat{U}_\mathrm{AD}$ operator~\cite{Ashida2021PRL} is no longer a rigorous double shift operator since it now has explicit $\hat{\bf x}$ dependence. Beyond the LWA, $\hat{U}_\mathrm{AD}$ now takes the form
\begin{equation}
    \hat{U}_\mathrm{AD} = \exp \Bigg[ -\frac{i}{\hbar} \sum_{\alpha} \boldsymbol{\xi}_{\alpha} \cdot \hat{\bf p} \, \hat{\Tilde{p}}_\alpha(\hat{\bf x}) \Bigg].
\end{equation}
With this, $\hat{U}_\mathrm{AD}$ is no longer $\hat{\bf x}$-independent, making it no longer rigorously behave as a double-shift operator. However, for states where $|{\bf k} \cdot \boldsymbol{\xi}| \ll 1$, this can be accurately approximated as a double shift operator~\cite{Ashida2021PRL}. In other words, we partially restore the LWA but now instead of claiming that the field is spatially invariant across the entire matter system, we make a less restrictive approximation that the $\hat{\bf x}$ dependence of the field is varying slowly enough such that it is negligible over the shift performed by the photonic DOF. In this manner, we explicitly make the approximation
\begin{align} \label{eq:lwa_ad}
    \hat{U}_\mathrm{AD}^\dagger \hat{\bf p} ~ \hat{U}_\mathrm{AD} &= {\hat{\bf p}} \big[ 1 + \sum_\beta \xi_\beta \cdot {\bf k}_\beta \omega_\beta \hat{\Tilde{q}}_\beta(\hat{\bf x}) + \cdots \big] \\
    & \approx {\hat{\bf p}}. \notag
\end{align}
Since $\boldsymbol{\xi}_\alpha$ is upper-bounded (see Fig.~\ref{FIG:MAIN_Xi_gc__loglog_Demler}a and Eq.~\ref{eq:xi_j_alpha}), for both zero coupling and arbitrarily high coupling, this approximation becomes exact, yielding the AD Hamiltonian beyond the LWA as
\begin{equation}\label{HAD_BLWA}
    \hat{H}_\mathrm{AD} = \frac{\hat{\bf p}^2}{2m_\mathrm{eff}}+\hat{V}({\hat{\bf x}} + \sum_\alpha {\xi}_\alpha \hat{\Tilde{p}}_\alpha(\hat{\bf x})) + \sum_\alpha \hbar \Omega_\alpha \Big(\hat{\Tilde{b}}^\dagger_\alpha(\hat{\bf x}) \hat{\Tilde{b}}_\alpha(\hat{\bf x}) + \frac{1}{2} \Big),
\end{equation}
where now $\hat{\Tilde{p}}_\alpha(\hat{\bf x})$ and $\hat{\Tilde{b}}_\alpha(\hat{\bf x})$ explicitly depend on the matter coordinate $\hat{\bf x}$. That is, every $\hat{a}_{\{ \lambda, {\bf k}_\beta \} }$ now has an additional $e^{i {\bf k}_\beta \cdot \hat{\bf x}}$ phase term associated with it. Note that after the normal mode transformation $\hat{\Tilde{b}}_\alpha(\hat{\bf x})$ and $\hat{\Tilde{b}}_\alpha^\dagger(\hat{\bf x})$ are defined in terms of $\hat{\Tilde{p}}_\alpha(\hat{\bf x})$ and $\hat{\Tilde{q}}_\alpha(\hat{\bf x})$ (as in Eq.~\ref{eq:b_a}) as follows
\begin{subequations} \label{eq:b_a_blwa}
    \begin{align}
    \hat{\Tilde{b}}_\alpha(\hat{\bf x}) &= \sqrt{\frac{\Omega_\alpha}{2 \hbar}} \hat{\Tilde{q}}_\alpha(\hat{\bf x}) + i \sqrt{\frac{1}{2 \hbar \Omega_\alpha}} \hat{\Tilde{p}}_\alpha(\hat{\bf x}) \\
    \hat{\Tilde{b}}_\alpha^\dagger(\hat{\bf x}) &= \sqrt{\frac{\Omega_\alpha}{2 \hbar}} \hat{\Tilde{q}}_\alpha(\hat{\bf x}) - i \sqrt{\frac{1}{2 \hbar \Omega_\alpha}} \hat{\Tilde{p}}_\alpha(\hat{\bf x})
    \end{align}
\end{subequations}

Recall that due to the $U(1)$ symmetry of QED, each photonic DOF is invariant under phase rotations (with the generator of the $\beta_\mathrm{th}$ mode defined as $\hat{a}^\dagger_\beta \hat{a}_\beta$) and the matter DOF is invariant upon a momentum boost (with the generator $\hat{\bf x}$). As such, we can now define a new operator that simultaneously performs a phase rotation on the $\beta_\mathrm{th}$ photonic mode and a boost on the electronic momentum (in the single-particle picture), expressed as
\begin{equation}\label{EQ:U_PHI}
    \hat{U}_{\phi_\beta} = \exp ({- i {\bf k}_\beta \cdot \hat{\bf x} \, \hat{a}^\dagger_\beta \hat{a}_\beta}),
\end{equation}
where ${\bf k}_\beta$ corresponds to the photonic wavevector of the $\beta_\mathrm{th}$ mode. This unitary transformation adds a $e^{-i \bf k_\beta \hat{\bf x}}$ phase to the $\hat{a}_\beta$ operator such that for any operator of the form $\hat{O}(\hat{a}_\beta e^{i {\bf k}_\beta \cdot \hat{\bf x}},\hat{a}_\beta^\dagger e^{-i {\bf k}_\beta \cdot \hat{\bf x}})$, transforming it would yield $\hat{U}_{\phi_\beta}^\dagger \hat{O}(\hat{a}_\beta e^{i {\bf k}_\beta \cdot \hat{\bf x}},\hat{a}_\beta^\dagger e^{-i {\bf k}_\beta \cdot \hat{\bf x}}) \hat{U}_{\phi_\beta} = \hat{O}(\hat{a}_\beta,\hat{a}_\beta^\dagger)$. Additionally, this operator boosts the matter momentum, so $\hat{U}_{\phi_\beta}^\dagger \hat{\bf p} \hat{U}_{\phi_\beta} = \hat{\bf p} - \hbar {\bf k}_\beta  \hat{a}^\dagger_\beta \hat{a}_\beta$.

Since $[\hat{U}_{\phi_\beta},\hat{U}_{\phi_{\beta'}}] = 0$, we can then write the phase rotation operator that eliminates the spatial variation of all modes as
\begin{equation}
    \hat{U}_{\phi} = \prod_\beta \hat{U}_{\phi_\beta} = \exp ({- i  \hat{\bf x} \cdot \Big(\sum_\beta {\bf k}_\beta \, \hat{a}^\dagger_\beta \hat{a}_\beta \Big) }),
\end{equation}
which then has the properties
\begin{subequations} \label{eq:U_phi_prop}
    \begin{gather}
        \hat{U}_{\phi }^\dagger \hat{O} \big(\{ \hat{a}_\beta e^{i {\bf k}_\beta \cdot \hat{\bf x}} \}, \{ \hat{a}_\beta^\dagger e^{-i {\bf k}_\beta \cdot \hat{\bf x}} \} \big) \hat{U}_{\phi } = \hat{O} \big(\{\hat{a}_\beta \}, \{ \hat{a}_\beta^\dagger \} \big)\\
        \hat{U}_{\phi }^\dagger \hat{\bf p} \, \hat{U}_{\phi } = \hat{\bf p} - \sum_\beta \hbar {\bf k}_\beta  \hat{a}^\dagger_\beta \hat{a}_\beta. \label{eq:u_phi_boost}
    \end{gather}
\end{subequations}
Since $\hat{\Tilde{q}}_\alpha, \hat{\Tilde{p}}_\alpha, \hat{\Tilde{b}}_\alpha,$ and $\hat{\Tilde{b}}^\dagger_\alpha$ are all functions of $\{ \hat{a}_\beta e^{i {\bf k}_\beta \cdot \hat{\bf x}} \}$ and $\{ \hat{a}_\beta^\dagger e^{-i {\bf k}_\beta \cdot \hat{\bf x}} \}$, this means that we can use $\hat{U}_{\phi}$ to remove the $\hat{\bf x}$ dependence of these operators, making the problem mathematically similar to the case with the LWA but under a smaller approximation. 

The properties of $\hat{U}_{\phi}$ shown in Eq.~\ref{eq:U_phi_prop}, allow us to transform $\hat{U}_{\phi}^\dagger \hat{H}_\mathrm{AD} \hat{U}_{\phi}$ as
\begin{align}
    \hat{U}_{\phi}^\dagger \hat{H}_\mathrm{AD} \hat{U}_{\phi} =&~ \frac{ \Big( \hat{\bf p} - \sum_\beta \hbar {\bf k}_\beta  \hat{a}^\dagger_\beta \hat{a}_\beta \Big)^2}{2m_\mathrm{eff}}+\hat{V} \Big( {\hat{\bf x}} + \sum_\alpha {\xi}_\alpha \hat{ {p}}_\alpha \Big)  \nonumber \\
    &+ \sum_\alpha \hbar \Omega_\alpha \Big(\hat{ {b}}^\dagger_\alpha \hat{ {b}}_\alpha + \frac{1}{2} \Big),
\end{align}
which is identical to the $\hat{H}_\mathrm{AD}$ under the LWA except for the boost of $- \sum_\beta \hbar {\bf k}_\beta  \hat{a}^\dagger_\beta \hat{a}_\beta$ to the momentum. Upon examination, this necessity of an additional boost can be thought of as a restatement of the conservation of momentum between the photonic and electronic degrees of freedom. For a photon with the momentum $\hbar {\bf k}$ to be created, the electron loses $\hbar {\bf k}$ momentum. In other words, the long-wavelength approximation in Eq.~\ref{eq:A_BLWA} {\it violates} the conservation of momentum between the photonic and electronic DOF. In fact, this conservation of momentum between light and matter is destroyed even when we use $\hat{\bf A} ( \hat{\bf x} ) \to \hat{\bf A} ( {\bf x} )$, {\it e.g.}, in the case of the multicenter PZW Hamiltonian~\cite{Li2020PRB, Mandal2023CR, Mandal2023NL}. This is because the operator nature of $\hat{\bf x}$ creates the momentum boost of Eq.~\ref{eq:u_phi_boost}, and by replacing it with its value $\bf x$, one no longer has the matter momentum shift, thus violating the conservation of momentum for the matter-photon hybrid system. As QED is gauge-independent, this conservation of momentum can also be seen in the Coulomb gauge Hamiltonian in Eq.~\ref{eq:h_pa_blwa} via $\hat{\bf A} ( \hat{\bf x} )$.

Now by following the same strategy of applying $\hat{U}_{\pi/2}$ (Eq.~\ref{EQ:Unitary_p_to_q_rotation}) and inserting identity in $K$-space (Eq.~\ref{eq:iden_k}) as is done before, the RAD Hamiltonian beyond the long-wavelength approximation for a single electron coupled to many modes then becomes
\begin{align} \label{eq:H_RAD_1M_BLWA}
    \hat{H}_\mathrm{RAD}^{[1][M]} =&~ \int d {\bf K} \,\, \frac{1}{2 m_\mathrm{eff}}
    \Big|\hbar {\bf K} \ketbra{{\bf K}}{{\bf K}} - \sum_\beta \hbar {\bf k}_\beta  \hat{a}^\dagger_\beta \hat{a}_\beta \Big|^2\\
    &+ \int d {\bf K} \,\, d {\bf K}'  ~ \, e^{ -{i {\bf K}' \cdot \sum_\alpha^M {\boldsymbol{\xi}}_{\alpha} \hat{q}_\alpha} } \mathcal{V}({\bf K}') \ketbra{{\bf K}}{{\bf K} +{\bf K}'} \nonumber  \\
    &+ \sum_\alpha^M \hbar \Omega_\alpha (\hat{b}_\alpha^\dagger \hat{b}_\alpha + \frac{1}{2}), \nonumber
\end{align}
where $\sum_\beta \hbar {\bf k}_\beta  \hat{a}^\dagger_\beta \hat{a}_\beta$ can be rewritten in terms of $\{ \hat{b}_\alpha, \hat{b}_\alpha^\dagger \}$ using the normal mode transformation matrix $\tensor{\hat{o}}$ (see Appendix~\ref{app:normal_modes}). As the last two lines of Eq.~\ref{eq:H_RAD_1M_BLWA} are identical to the corresponding terms in Eq.~\ref{eq:h_rad-1par}, the extension to periodic systems follows identically to before, yielding
\begin{align}\label{eq:h_rad_period_blwa}
    &\hat{H}_{\mathrm{RAD}} = \sum_\alpha^M \hbar \Omega_\alpha (\hat{b}_\alpha^\dagger \hat{b}_\alpha + \frac{1}{2}) \\
    &+ \sum_{\boldsymbol{\kappa}} \int_\mathrm{1BZ} d {\bf k} \,\, \frac{1}{2 m_\mathrm{eff}}
    \Big|\hbar ({\bf k} + \boldsymbol{\kappa}) \ketbra{{\bf k} + \boldsymbol{\kappa}}{{\bf k} + \boldsymbol{\kappa}} - \sum_\beta \hbar {\bf k}_\beta  \hat{a}^\dagger_\beta \hat{a}_\beta \Big|^2 \nonumber \\
    &+ \sum_{\boldsymbol{\kappa}, \boldsymbol{\kappa}'} \int_\mathrm{1BZ} d {\bf k} \,\, e^{ -{i \boldsymbol{\kappa}' \cdot \sum_\alpha^M {\boldsymbol{\xi}}_{\alpha} \hat{q}_\alpha} } v(\boldsymbol{\kappa}') \ketbra{{\bf k}+\boldsymbol{\kappa}}{{\bf k} + \boldsymbol{\kappa} +\boldsymbol{\kappa}'} \nonumber  , 
\end{align}
where the integral of $\bf k$ goes over the $1_\mathrm{st}$ Brillouin zone and $\boldsymbol{\kappa}$ and $\boldsymbol{\kappa}'$ are reciprocal lattice vectors.

It can be observed that when the electronic system couples to the electromagnetic field, it breaks the translational invariance (cf. Bloch's Theorem) along the electronic coordinate due to the spatial variations of the field (see Eq.~\ref{eq:A_BLWA}) not following the periodicity of the matter. However, by applying $\hat{U}_\phi$ (taking $\hat{\bf p} \to \hat{\bf p} + \sum_\beta \hbar {\bf k}_\beta \hat{a}^\dagger_\beta \hat{a}_\beta$), we regain translational invariance in this boosted {\it polaritonic} space as shown in Eq.~\ref{eq:h_rad_period_blwa}. Following a similar analysis as in Eqs.~\ref{eq:p_k}-\ref{eq:h_rad_k_gen} but without the explicit restriction of ${\bf k} = {\bf k}_\beta$, we then project Eq.~\ref{eq:h_rad_period_blwa} in terms of both ${\bf k}$ and ${\bf k}_\beta$. This allows us to generate dispersion plots for a single particle beyond the LWA, where we parameterize the Hamiltonian as
\begin{align}\label{eq:h_rad_period_blwa_k}
    &\hat{H}_{\mathrm{RAD}} ({\bf k, {\bf k_\beta}}) = \hbar \Omega_{\bf k_\beta} (\hat{b}_{\bf k_\beta}^\dagger \hat{b}_{\bf k_\beta} + \frac{1}{2}) \\
    &+ \sum_{\boldsymbol{\kappa}} \frac{1}{2 m_\mathrm{eff}}
    \Big|\hbar ({\bf k} + \boldsymbol{\kappa}) \ketbra{{\bf k} + \boldsymbol{\kappa}}{{\bf k} + \boldsymbol{\kappa}} - \hbar {\bf k}_\beta  \hat{a}^\dagger_\beta \hat{a}_\beta \Big|^2 \nonumber \\
    &+ \sum_{\boldsymbol{\kappa}, \boldsymbol{\kappa}'} e^{ -{i \boldsymbol{\kappa}' \cdot {\boldsymbol{\xi}}_{\bf k_\beta} \hat{q}_{\bf k_\beta}} } v(\boldsymbol{\kappa}') \ketbra{{\bf k}+\boldsymbol{\kappa}}{{\bf k} + \boldsymbol{\kappa} +\boldsymbol{\kappa}'} \nonumber . 
\end{align}
This is the final {\it key result} of this paper.

\begin{figure}
 \centering
 %%-------start of first figure----------
    \begin{minipage}[h]{\linewidth}
     \centering
     \includegraphics[width=0.94\linewidth]{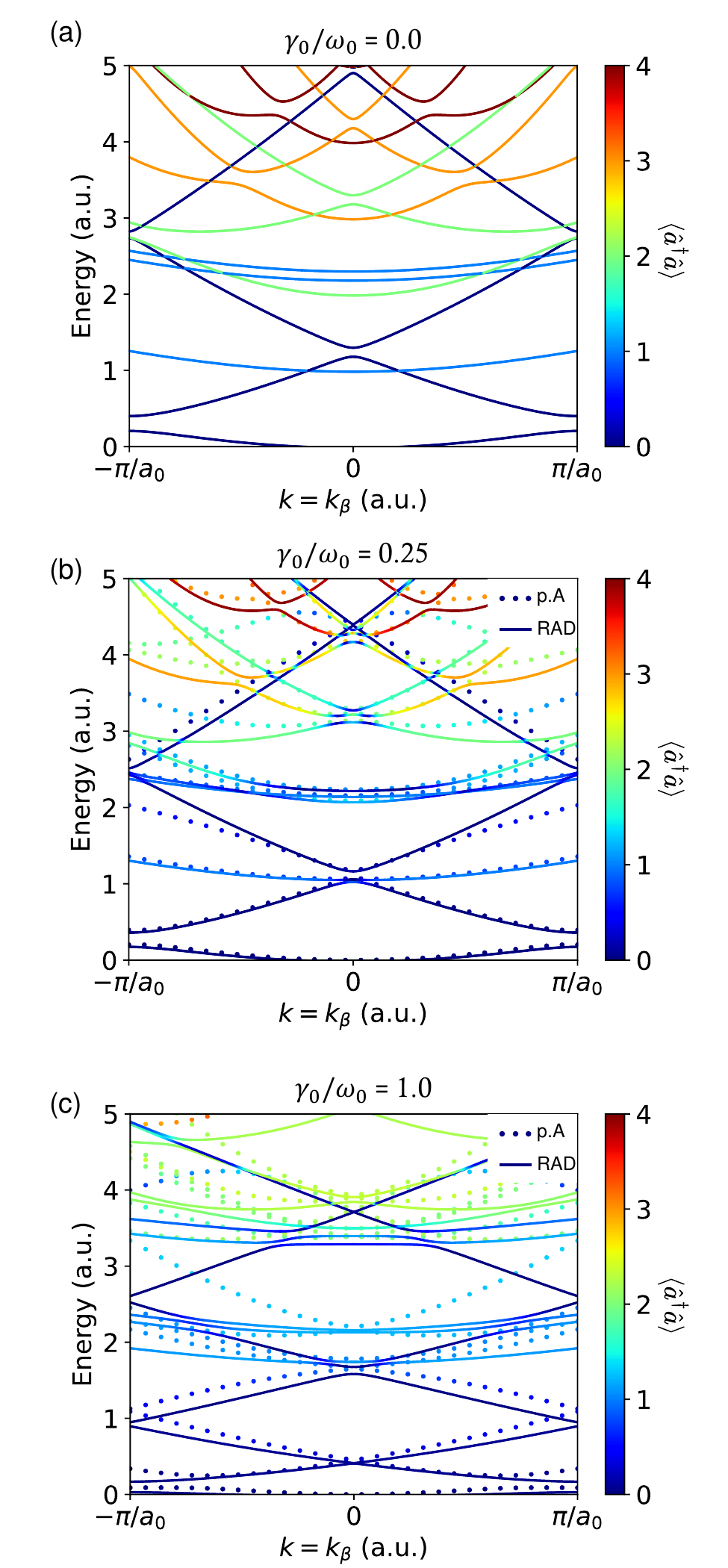}
    \end{minipage}%
   \caption{Single electron in a periodic modified Coulomb potential coupled to many longitudinal cavity modes beyond the LWA for both the RAD Hamiltonian (solid lines) and the Coulomb gauge Hamiltonian (dotted lines). (a) Zero-coupling case for which both RAD and p$\cdot$A are exact. (b) Intermediate coupling case ($\gamma_0 / \omega_0 = 0.25$) where the approximation in Eq.~\ref{eq:lwa_ad} starts to make RAD no longer perfectly match the Coulomb gauge. (c) Coupling of $\gamma_0 / \omega_0 = 1.0$, where $\xi_{k_\beta}$ maximizes, representing the most challenging case for RAD. Note that for coupling strengths beyond $\gamma_0 / \omega_0 = 1.0$, RAD will get increasingly more accurate and converge faster, whereas the Coulomb gauge results will require increasingly more Fock states and matter bands to converge.}
   \label{fig:blwa_erf_pot}
\end{figure}

Fig.~\ref{fig:blwa_erf_pot} plots the band dispersions of the 1D modified Coulomb potential for the cross-section $k = k_\beta$. We begin our discussion with the simplest case of zero coupling, as shown in Fig.~\ref{fig:blwa_erf_pot}a. At first glance, this plot differs greatly from the types of plots in Fig.~\ref{fig:erf_pot}. We note that only the diagonal matrix elements of $\hat{H}_{\mathrm{RAD}} ({\bf k, {\bf k_\beta}})$ vary with $k$ and $k_\beta$ for zero coupling, taking the form
\begin{equation} \label{eq:h_rad_diag_0}
    \bra{k + \kappa, n_\beta} \hat{H}_{\mathrm{RAD}}  \ket{k + \kappa, n_\beta} = \hbar \omega_\beta n + \frac{\hbar^2}{m} \big| \kappa + k - n k_\beta \big|^2,
\end{equation}
where for simplicity we subtract out any zero-point energies. At zero coupling, $\hat{b}_{k_\beta} = \hat{a}_\beta$ and $\Omega_{k_\beta} = \omega_\beta$ (\textit{i.e.}, the RAD and Coulomb representations of the photon operators are identical), allowing us to equivalently understand the $n_\mathrm{th}$ Fock state as $\ket{n_\beta} = \frac{1}{\sqrt{n!}}(\hat{a}_\beta^\dagger)^n \ket{0}$ or $\ket{n_\beta} = \frac{1}{\sqrt{n!}} (\hat{b}_\beta^\dagger)^n \ket{0}$. Using Eq.~\ref{eq:h_rad_diag_0} we can make sense of Fig.~\ref{fig:blwa_erf_pot}a, where we plot the cross-section of the polaritonic dispersion for $k = k_\beta$, by focusing on the bands of a given photon number. Without any light-matter coupling, the zero-photon bands in Fig.~\ref{fig:blwa_erf_pot}a ($n=0$, dark blue curves) exactly follow the bare matter band dispersion. Then, for the one-photon ($n=1$) bands, $k - n k_\beta = 0$, making the bands only have the single-photon dispersion shifted by the matter band energies at the $\Gamma$-point, creating three visible light blue parabolic curves in Fig.~\ref{fig:blwa_erf_pot}a. For the $n > 1$ bands, $k - n k_\beta = (1-n) k$. This replicates the bare matter bands shifted up by the energy $\hbar c k_\beta$; however, the matter Brillouin zones shrink by a factor of $n-1$ such that for $n=3$, two of the bare matter Brillouin zones are squeezed into the system's first Brillouin zone. 

We would like to emphasize that the $k$ plotted in the dispersion plots in Fig.~\ref{fig:blwa_erf_pot} is no longer the eigenvalue of the canonical momentum, as we have boosted the momentum such that $\hat{\bf p} \to \hat{\bf p} + \sum_\beta \hbar {\bf k}_\beta \hat{a}^\dagger_\beta \hat{a}_\beta$. As such, $k$ is no longer exactly the matter lattice wavevector, since the light-matter system is no longer translationally invariant by the period of the lattice. Instead, $k$ is akin to a ``polaritonic wavevector'' in that two states with the same $k$ have the same total momentum on the system level. This quantity takes advantage of the symmetry of the system, so, unlike the matter lattice wavevector, $\hat{H}_\mathrm{RAD}$ is block diagonal in $k$. This feature of the RAD representation beyond the LWA allows us to calculate realistic ``polaritonic dispersions'' and visualize the light-matter hybridized bands.

Additionally, these polaritonic dispersions now allow for an easy visualization of how transitions between polaritonic states change the momentum and energy of the system. For example, the crossing of two bands represents a degenerate point in both energy and momentum. These types of plots are reminiscent of the type of graphs used to visualize other light-matter interactions such as Brillouin and Raman scattering of photons.

While the prior analysis is numerically exact for the zero-coupling case, for nonzero coupling, we are still making a form of the LWA as is stated in Eq~\ref{eq:lwa_ad}. The natural question is the validity of such an approximation. To benchmark the RAD result, we must calculate the polaritonic dispersion plots for the exact Coulomb gauge Hamiltonian. By going beyond the LWA, we reintroduced in Eq.~\ref{eq:h_pa_blwa} the $e^{\pm i {\bf k}_\beta \cdot \hat{\bf x}}$ terms in the vector potential, $\hat{\bf A}$. These terms make $\hat{H}_\mathrm{p \cdot A}$ no longer block diagonal in $\bf k$. This can be explicitly seen by transforming $\hat{H}_\mathrm{p \cdot A}$ into reciprocal space.
\begin{align} \label{h_pa_k}
\hat{H}_\mathrm{p \cdot A}& = \int d{\bf K} \,\, \frac{1}{2m}  \bigg( \hbar^2 |{\bf K}|^2 \ketbra{{\bf K}}{{\bf K}} + \big(\sum_{\beta} z |{\bf A} |  ( \hat{a}_{\beta}^\dagger \hat{B}_\beta + \hat{a}_{\beta} \hat{B}_\beta^\dagger) \big)^2 \nonumber \\ 
&-  \sum_{\beta} \hbar z {\bf K} \cdot {\bf A}_{{\beta}} \Big\{ \ketbra{{\bf K}}{{\bf K}}, ( \hat{a}_{\beta}^\dagger \hat{B}_\beta + \hat{a}_{\beta} \hat{B}_\beta^\dagger) \Big\} \bigg) \nonumber \\ 
&+\int d {\bf K} \,\, d {\bf K}'  ~ \, \mathcal{V}({\bf K}') \ketbra{{\bf K}}{{\bf K} +{\bf K}'} + \sum_\beta \hbar\omega_\beta \Big(\hat{a}^{\dagger}_\beta \hat{a}_\beta +\frac{1}{2}\Big),
\end{align}
where we have introduced the matter momentum boost operators, $\hat{B}_\beta = \int d{\bf K'} \ketbra{\bf K' - k_\beta}{\bf K'}$ and $\hat{B}_\beta^\dagger = \int d{\bf K'} \ketbra{\bf K' + k_\beta}{\bf K'}$, which come from sandwiching the $e^{\pm i {\bf k}_\beta \cdot \hat{\bf x}}$ terms with identities of the form of Eq.~\ref{eq:iden_k}. These $\hat{B}_\beta$ and $\hat{B}_\beta^\dagger$ terms are clearly not block diagonal in $\bf k$ even for periodic lattices, since, in principle, $k_\beta$ is quasi-continuous. As such, these coupling terms between different matter $\bf k$-points break the original Bloch's theorem along the electronic coordinate. However, this way of writing $\hat{\bf A} = \sum_\beta {\bf A}_{{\beta}} (\hat{a}_{\beta}^\dagger \hat{B}_\beta + \hat{a}_{\beta} \hat{B}_\beta^\dagger)$ also shows how the $e^{\pm i {\bf k}_\beta \cdot \hat{\bf x}}$ terms lead to the {\it conservation of momentum} between the photonic and electronic DOFs: for each creation of a photon with momentum $\hbar \bf k_\beta$ by $\hat{a}_\beta^\dagger$ the same amount of momentum is boosted away from the electron by $\hat{B}_\beta$ and vice versa for $\hat{a}_\beta$ and $\hat{B}_\beta^\dagger$. Eq.~\ref{h_pa_k} thus reinforces the necessity of including the $e^{\pm i {\bf k}_\beta \cdot \hat{\bf x}}$ terms and simultaneously how such terms destroy the electronic coordinate's translational invariance.

It may be tempting to try to resolve this difficulty by simply replacing $e^{\pm i {\bf k}_\beta \cdot \hat{\bf x}}$ with $e^{\pm i {\bf k}_\beta \cdot {\bf x}_u}$, where ${\bf x}_u$ is the location of the $u_\mathrm{th}$ lattice site, rewriting the Hamiltonian in the site basis and making the approximation that the field varies slowly across the lattice unit cell~\cite{Dmytruk2021PRB,Li2020PRB,Mandal2023NL}. This approximation has been thoroughly investigated in the context of the multicenter PZW transformation, where the polaritonic Hamiltonian is expressed in the dipole gauge~\cite{Li2020PRB,Mandal2023NL} (see Sec.~2.6.1 in Ref.~\cite{Mandal2023CR} for details). The resulting Hamiltonian does, in fact, satisfy Bloch's theorem since there no longer are any $\hat{B}_\beta$ {\it operators} in reciprocal space, but it still violates the conservation of momentum.
% Additionally, making this type of approximation adds an additional restraint that $\bf k = k_\beta$. 

Nevertheless, we can apply a strategy from the derivation of $\hat{H}_\mathrm{RAD}$ from Eq.~\ref{eq:H_RAD_1M_BLWA} to restore both Bloch's theorem and the conservation of momentum without making any approximations to Eq.~\ref{h_pa_k}. Intuitively, we know from Eq.~\ref{h_pa_k} that the $\hat{\bf x}$-dependence in $\hat{\bf A} (\hat{\bf x})$ acts as a momentum boost for the electron that balances out the momentum change from the creation/annihilation of a photon. By absorbing the photon and electronic DOFs' momenta into a system-wide ``polariton'' momentum, these boost operators would no longer explicitly appear. This is reminiscent of the ``conservation of crystal momentum'' argument used for electron-phonon interactions, where the electron and phonon momenta are grouped into a {\it total crystal momentum} (See Appendix M in Ref.~\citenum{AshcroftMermin}). This change of variables, to a total ``polaritonic'' momentum, allows us to reformulate this exact Hamiltonian using Bloch's theorem. Thus, by transforming the exact Coulomb gauge Hamiltonian, $\hat{H}_\mathrm{p \cdot A}$, by $\hat{U}_\phi$ (see Eq.~\ref{EQ:U_PHI}) we can write an exact Hamiltonian that is block diagonal in $\bf k$ and conserves momentum as
\begin{align} \label{h_pa_blwa_k}
    \hat{U}_\phi^\dagger \hat{H}_\mathrm{p \cdot A} \hat{U}_\phi =&~ \frac{1}{2m} \Big(\hat{\bf p}-{z}\sum_{\beta} {\bf A}_{{\beta}} ( \hat{a}_{\beta}^\dagger  + \hat{a}_{\beta} ) -\sum_\beta \hbar {\bf k}_\beta \hat{a}^\dagger_\beta \hat{a}_\beta \Big)^2 \nonumber \\ 
    &+\hat{V}(\hat{\bf x})+ \sum_\beta \hbar\omega_\beta \Big(\hat{a}^{\dagger}_\beta \hat{a}_\beta +\frac{1}{2}\Big),
\end{align}
where as with $\hat{H}_\mathrm{RAD}$ in Eq.~\ref{eq:H_RAD_1M_BLWA} we transformed $\hat{\bf p} \to \hat{\bf p} + \sum_\beta \hbar {\bf k}_\beta \hat{a}^\dagger_\beta \hat{a}_\beta$. This expression in Eq.~\ref{h_pa_blwa_k} is exact and for our simple system can be directly calculated. To do so, we parameterize this expression by $\bf k$ and $\bf k_\beta$ as
\begin{align}\label{EQ:H_pA_PHI_k_kbeta}
    &\hat{U}_\phi^\dagger \hat{H}_\mathrm{p \cdot A} \hat{U}_\phi ({\bf k}, {\bf k_\beta}) = \hbar\omega_\beta \Big(\hat{a}^{\dagger}_\beta \hat{a}_\beta +\frac{1}{2}\Big) \\ 
    &~~~~~~~~~~~~~~+ \sum_{\boldsymbol{\kappa}} \frac{1}{2 m}
    \Big(\hbar ({\bf k} + \boldsymbol{\kappa}) \ketbra{{\bf k} + \boldsymbol{\kappa}}{{\bf k} + \boldsymbol{\kappa}} -{z}{\bf A}_{{\beta}} ( \hat{a}_{\beta}^\dagger  + \hat{a}_{\beta} )  \nonumber \\
    &~~~~~~~~~~~~~~- \hbar {\bf k}_\beta  \hat{a}^\dagger_\beta \hat{a}_\beta \Big)^2 + \sum_{\boldsymbol{\kappa}, \boldsymbol{\kappa}'}  v(\boldsymbol{\kappa}') \ketbra{{\bf k}+\boldsymbol{\kappa}}{{\bf k} + \boldsymbol{\kappa} +\boldsymbol{\kappa}'}, \nonumber
\end{align}
where we can numerically calculate the polaritonic dispersion plots directly as we did with $\hat{H}_\mathrm{RAD}({\bf k}, {\bf k_\beta})$. We use this Hamiltonian as a benchmark to check the validity of our approximation Eq.~\ref{eq:lwa_ad}. The convergence of this $\mathrm{p \cdot A}$ Hamiltonian is slower than $\hat{H}_\mathrm{RAD}({\bf k}, {\bf k_\beta})$ (Eq.~\ref{eq:h_rad_period_blwa_k}), requiring 11 bands ($\kappa$ values) and 14 $\mathrm{p \cdot A}$ Fock states to converge, whereas the RAD Hamiltonian only required 7 bands ($\kappa$ values) and 5 RAD Fock states to converge the results plotted, which is especially remarkable due to there being 5 bands plotted in Fig.~\ref{fig:blwa_erf_pot}c. Notably, this reduction in dimension $d$ (from 154 for $\mathrm{p \cdot A}$ to 35 for RAD) allows for a factor of $\sim$85 speedup for the direct matrix diagonalization given $\mathcal{O}(d^3)$ scaling.

Figs.~\ref{fig:blwa_erf_pot}b,c show two different coupling strengths with the results of RAD and the exact Coulomb gauge Hamiltonian. In Fig.~\ref{fig:blwa_erf_pot}b, the coupling strength $\gamma_0 / \omega_0 = 0.25$ is set at an intermediate value such that $\xi_{k_\beta}$ is large but not maximized (see Fig.~\ref{FIG:MAIN_Xi_gc__loglog_Demler}a), while in Fig.~\ref{fig:blwa_erf_pot}c, the coupling strength $\gamma_0 / \omega_0 = 1.0$ is set to the maximal value of $\xi_{k_\beta}$ and thus represents the most challenging case for the RAD Hamiltonian. From the zero-coupling analysis, the RAD results in Fig.~\ref{fig:blwa_erf_pot}b make intuitive sense with Rabi splitting at the crossing of bands (most clearly seen near $k = k_\beta \sim 0$ and energy $E \sim 3.0$ with the mixing of the green $n=2$ and orange $n=3$ curves). Additionally, in this regime, the RAD results generally match the Coulomb gauge results, with the most obvious disagreements occurring at large matter $k$, as expected from our assumption of $\xi_\mathrm{max} \cdot k_\mathrm{max} << 1$. 

\begin{figure}
 \centering
 %%-------start of first figure----------
    \begin{minipage}[h]{\linewidth}
     \centering
     \includegraphics[width=1.0\linewidth]{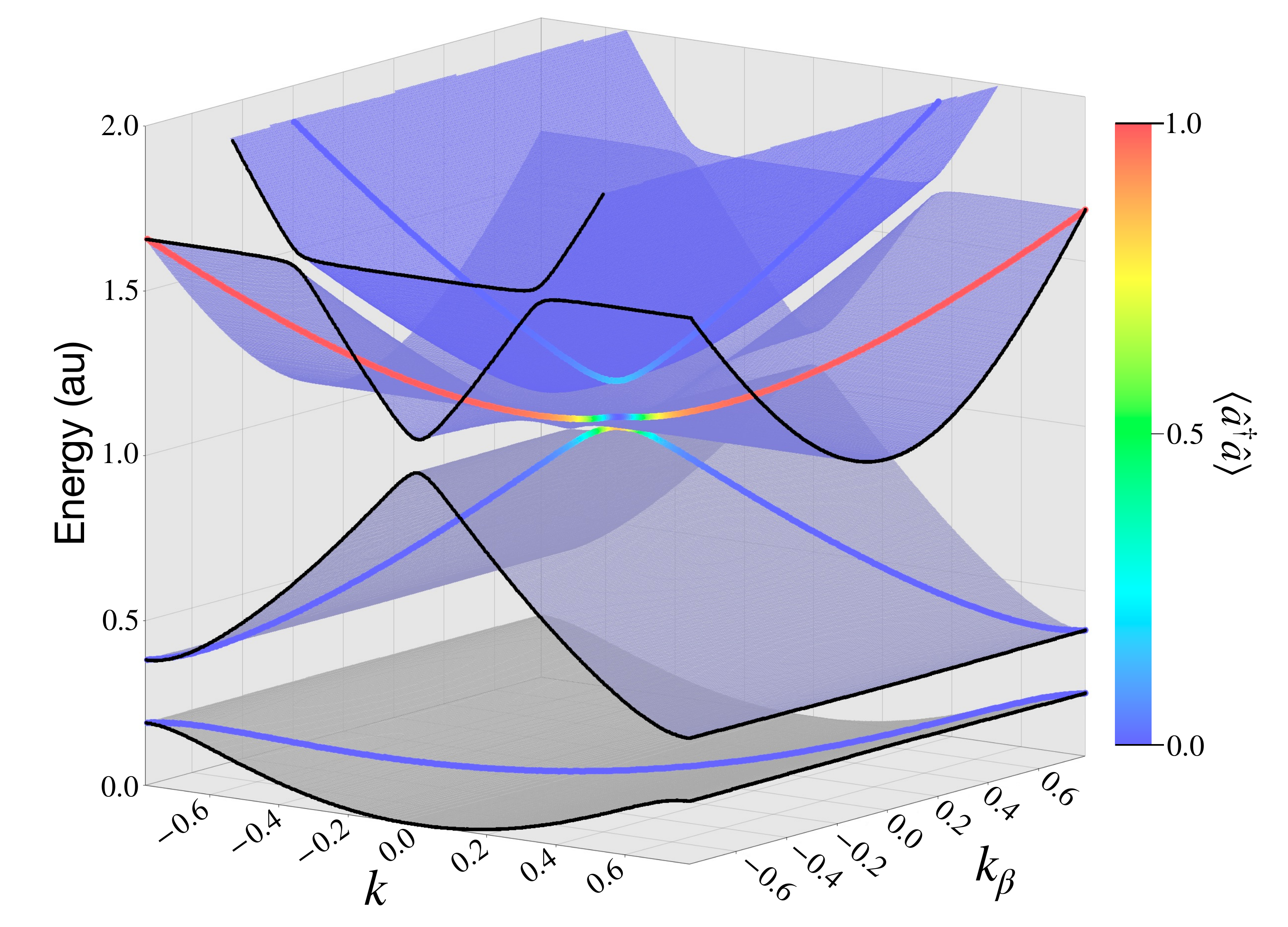}
    \end{minipage}%
   \caption{Two-dimensional polariton dispersion relation for the first four polariton bands for a coupling of $\gamma_0/\omega_0 = 0.25$, where surfaces are colored by band number. The $k = k_\beta$ cross-section is also plotted on these surfaces weighted by the photonic character with blue being electronic and red being photonic. Additionally, the edges of the plot are accented in black to help parse the 3D surfaces.}
   \label{fig:blwa_2d}
\end{figure}

Fig.~\ref{fig:blwa_erf_pot}c then goes on to test RAD for the worst coupling strength for our approximation, $\gamma_0 / \omega_0 = 1$, the maxima of $\xi_{k_\beta}$. Even in this case, the characteristic properties that the RAD Hamiltonian predicts are still valid. The single photon band is blue-shifted, and the matter bands collapse closer together due to the increase of the effective mass while also flattening due to the diminishing presence of the kinetic energy term (due to the effective mass). Since the only approximation in this theory is that of Eq.~\ref{eq:lwa_ad}, we know that as the coupling increases to $\gamma_0 / \omega_0 > 1$ the RAD results will get increasingly more accurate.

Fig.~\ref{fig:blwa_2d} presents the 2D dispersion relation of the polariton states along both $k$ and $k_\beta$. Note that the polariton dispersion curves from Figs.~\ref{fig:erf_pot} and \ref{fig:blwa_erf_pot} are only cross sections of the two-dimensional dispersions of both DOFs's momentum, $k$ and $k_\beta$, where we take the diagonal cross-section of $k = k_\beta$ (since we transformed $k \to k + k_\beta$, this is the cross-section through the matter $\Gamma$-point). This cross-section provides a convenient picture for transitions from the $\Gamma$-point (\textit{i.e.}, matter $k = 0$) but does not show all possible states of the hybrid system. Note that for flat dispersion bands (such as those for collections of noninteracting molecules in a Fabry-P\'erot cavity), only the matter $\Gamma$-point is optically bright, so this cross-section fully characterizes the system~\cite{Mandal2023NL}. However, to understand transitions from an electron with momentum $k_0$ for matter systems with a non-trivial band structure, a cross-section of $k - k_\beta = k_0$ would provide more intuition. To understand the full picture, the full 2D polariton dispersion plot is necessary. The $k = k_\beta$ cross-section is also drawn on top of the surfaces of Fig.~\ref{fig:blwa_2d} with the color indicating the photonic character (represented by $\langle \hat{a}^\dagger_\beta \hat{a}_\beta \rangle$) where blue represents purely electronic and red purely photonic. 

By going beyond the long-wavelength approximation, the conservation of momentum between light and matter is restored. This allows for more physically intuitive band dispersions, allowing one to understand the absorption and emission phenomena of such periodic systems without performing direct absorption or photoluminescence spectra.
 
\section{Conclusions}
In conclusion, we developed a new representation that can accurately and efficiently calculate the eigenenergies of polariton systems for arbitrarily strong coupling strengths.  The computational cost to calculate the eigenspectra using existing Hamiltonians (such as $\mathrm{p\cdot A}$ or $\mathrm{d\cdot E}$) scales unfavorably with increasing coupling strength. We began by reviewing the Asymptotically Decoupled (AD) Hamiltonian presented by Ashida et al. in Ref.~\citenum{Ashida2021PRL}. 

Sec.~\ref{sec:AD} generalized the key result of Ref.~\citenum{Ashida2021PRL} into a form for many interacting charged particles coupled to many photon modes (see Eq.~\ref{eqn:AD_ham}). This is accomplished via a normal mode transformation (Bogoliubov transform for a single cavity mode) followed by a many-particle and many-mode double-shift operator (Eq~\ref{eq:uad}). While this representation has a much better Fock state convergence than typical gauges (such as the dipole gauge Hamiltonian) and has an upper-bounded effective coupling parameter, the shift in the matter coordinates by the photonic momentum (see Eq.~\ref{eqn:AD_ham}) makes realistic calculations (or even more complicated model systems like that of Fig.~\ref{fig:erf_pot}) unfeasible.

To address this challenge, we introduced the Reciprocal Asymptotically Decoupled Hamiltonian (RAD) in Sec.~\ref{sec:RAD}. In particular, we applied a phase rotation unitary transformation (Eq.~\ref{EQ:Unitary_p_to_q_rotation}) and transformed it into reciprocal space, leveraging the Fourier Shift Theorem (Eq.~\ref{eq:FShift}). By doing so, the RAD Hamiltonian still holds the advantages of the AD Hamiltonian, but the matter coordinate is no longer shifted by the photonic momentum. Instead, the Fourier transform of the many-body potential is multiplied by a simple phase term of the form, $e^{-i \sum_{j,\alpha} {\bf K}_j \cdot {\boldsymbol{\xi}_{j,\alpha}} \hat{q}_\alpha}$ (in Eq.~\ref{Eq:V_RAD_shift}). This allows the eigenenergies to be calculated for any type of model or realistic single-particle potential. As discussed in Sec.~\ref{sec:RAD}, this RAD representation significantly outperforms the Pauli-Fierz Hamiltonian with Fock states and matter basis convergence, even with highly localized potentials such as the double-well potentials (See Fig.~\ref{FIG:MAIN_POTS_WFNS_DIPS}). The PF Hamiltonian still has the benefit of being able to diagonalize the matter system first, followed by a direct diagonalization of the light-matter Hamiltonian~\cite{weight_abQED_JPCL2023}. The RAD Hamiltonian, on the other hand, requires one to diagonalize all DOFs simultaneously without knowledge of the bare-matter states, and performing ab initio polariton simulations with the RAD Hamiltonian on realistic systems is a subject of future work.

As RAD is formulated in reciprocal space, its prime application is for periodic systems. Sec.~\ref{sec:period} applies the RAD Hamiltonian to the special case of $\hat{V}$ being periodic in nature. Assuming the long-wavelength approximation (LWA), by applying Bloch's Theorem to RAD (Eq.~\ref{V-pot-conv-3D}), and projecting the Hamiltonian to different $k$-points (see Eq.~\ref{eq:p_k}), polariton dispersion plots can be calculated. To go beyond the capabilities of the AD Hamiltonian, in Sec.~\ref{sec:1Derf} we used the RAD Hamiltonian to calculate the dispersion relations of a single electron in a 1D lattice of modified Coulomb potentials (See Fig.~\ref{fig:erf_pot}).

Finally, in Sec.~\ref{sec:BLWA} we take RAD beyond the long-wavelength approximation for a single particle coupled to many cavity modes, treating the electromagnetic field as spatially varying as a function of $\hat{\bf x}$. Doing so violates the conservation of momentum between the photonic and electronic degrees of freedom for the hybrid system, as well as Bloch's theorem. We resolve this issue in the single particle limit by introducing a new unitary transformation, $\hat{U}_\phi$, (Eq.~\ref{EQ:U_PHI}) that removes the explicit $\hat{\bf x}$-dependence of the field by grouping the photonic and electronic momenta into $\hat{\bf p}$.
This treatment preserves the conservation of momentum between the light and matter DOFs, allowing the calculation of physically relevant ``polaritonic dispersion'' curves that provide an intuitive understanding of absorption and emission processes. Additionally, in this section, we emphasize that one must be extremely careful in making any LWA, as it can violate the conservation of momentum between the light and matter DOFs. To benchmark this new method of going beyond the long-wavelength approximation, we also compared these results to the exact p$\cdot$A Hamiltonian transformed by $\hat{U}_\phi$.

This work opens many future directions in studying polariton physics. For example, this Hamiltonian can be immediately applied to any 1-electron \textit{ab initio} systems' model potentials. Additionally, this representation could be extended to include many-electron polariton systems. This work will enable investigations of periodic cavity QED systems and light-matter coupling in the ultra-strong and deep-strong coupling regimes~\cite{Kockum2019NRP}.

\section*{Acknowledgement}
This material is based upon work supported by the Air Force Office of Scientific Research under AFOSR Award No. FA9550-23-1-0438. M.T. appreciates the support from the National Science Foundation Graduate Research Fellowship Program under Grant No. DGE-1939268. Computing resources were provided by the Center for Integrated Research Computing (CIRC) at the University of Rochester. We appreciate valuable discussions with Arkajit Mandal, John Alejandro Montilla Ortega and Vishal Tiwari.

\appendix

\section{Normal Mode Analysis For the Quantized Field} \label{app:normal_modes}
As discussed in the main text, the standard Coulomb gauge Hamiltonian can be expressed in the form
\begin{align}\label{eq:Hc-app}
\hat{H}_\mathrm{p \cdot A}=&~ \hat{H}_\mathrm{M} -  \sum_{ j,\beta} \frac{z_j \hat{\bf p}_j \cdot {\bf A}_\beta}{m_j} \sqrt{\frac{2 \omega_\beta}{\hbar}}\hat{q}_\beta  \\
&+ \sum_{\beta,\beta'} \frac{1}{2} \Big[\hat{p}_\beta^2 \delta_{\beta,\beta'} + \big(\omega_\beta^2 \delta_{\beta,\beta'} + 2 \gamma_\beta \gamma_{\beta'}({\bf e}_\beta \cdot {\bf e}_{\beta'}) \big) \hat{q}_\beta \hat{q}_{\beta'} \Big], \nonumber
\end{align}
where we introduced a new mode-dependent coupling parameter,  
\begin{equation} \label{eq:g-app}
    \gamma_\beta = |\mathbf{A_\beta}| \sqrt{ \left( \frac{\omega_\beta}{\hbar} \right) \sum_j \frac{z_j^2}{m_j}},
\end{equation}
and we defined $\hat{p}_\beta$ and  $\hat{q}_\beta$ as,
\begin{subequations}
\begin{align}\label{Eq:q-beta-app}
\hat{q}_\beta = \sqrt{\frac{\hbar}{2\omega_\beta}}(\hat{a}^{\dagger}_\beta + \hat{a}_\beta)\\
\hat{p}_\beta = i\sqrt{\frac{\hbar\omega_\beta}{2}}(\hat{a}^{\dagger}_\beta - \hat{a}_\beta) \label{Eq:p-b-app}
\end{align}
\end{subequations}

To perform the normal mode analysis, it is convenient to define the vectors of operators, $\Vec{\hat{q}}_\mathrm{ph}$, $\Vec{\hat{p}}_\mathrm{ph}$, and $\Vec{\hat{\boldsymbol \zeta}}$, and second order tensor of operators, $\tensor{\hat{g}}$ as
\begin{subequations}
    \begin{gather}
        \Vec{\hat{q}}_\mathrm{ph}^\intercal = \begin{bmatrix}
            {\hat{q}}_{0}     &
            {\hat{q}}_{1}     &
            \cdots            &
            {\hat{q}}_{\beta} &
            \cdots
        \end{bmatrix}
         \\
         \Vec{\hat{p}}_\mathrm{ph}^\intercal = \begin{bmatrix}
            {\hat{p}}_{0}     &
            {\hat{p}}_{1}     &
            \cdots            &
            {\hat{p}}_{\beta} &
            \cdots
        \end{bmatrix}
         \\
         \Vec{\hat{\boldsymbol \zeta}}^\intercal = \begin{bmatrix}
            {\bf A}_{0} \sqrt{\omega_0} \otimes \hat{\mathds{1}}_\mathrm{ph} &
            {\bf A}_{1} \sqrt{\omega_1} \otimes \hat{\mathds{1}}_\mathrm{ph} &
            {\bf A}_{2} \sqrt{\omega_2} \otimes \hat{\mathds{1}}_\mathrm{ph} &
            \cdots
        \end{bmatrix}
         \\
        % \tensor{\hat{g}} = \begin{bmatrix}
        %     \omega^2_0 \gamma_0 \gamma_0 \otimes \hat{\mathds{1}}_\mathrm{ph} & \gamma_1 \gamma_0 \otimes \hat{\mathds{1}}_\mathrm{ph} & \gamma_2 \gamma_0 \otimes \hat{\mathds{1}}_\mathrm{ph}& \cdots \\
        %     \gamma_0 \gamma_1 \otimes \hat{\mathds{1}}_\mathrm{ph} & \omega^2_1 \gamma_1 \gamma_1 \otimes \hat{\mathds{1}}_\mathrm{ph} & \gamma_2 \gamma_1 \otimes \hat{\mathds{1}}_\mathrm{ph}& \cdots \\
        %     \gamma_0 \gamma_2 \otimes \hat{\mathds{1}}_\mathrm{ph} & \gamma_1 \gamma_2 \otimes \hat{\mathds{1}}_\mathrm{ph} & \omega^2_2 \gamma_2 \gamma_2 \otimes \hat{\mathds{1}}_\mathrm{ph} & \cdots \\
        %     \vdots & \vdots & \vdots & \ddots
        % \end{bmatrix}
        \tensor{\hat{g}} = \begin{bmatrix}
            (\omega^2_0 + 2 \gamma_0^2)\otimes \hat{\mathds{1}}_\mathrm{ph} & (2 \gamma_1 \gamma_0 {\bf e}_0 \cdot {\bf e}_{1} ) \otimes \hat{\mathds{1}}_\mathrm{ph} & \cdots \\
            (2 \gamma_0 \gamma_1 {\bf e}_1 \cdot {\bf e}_{0}) \otimes \hat{\mathds{1}}_\mathrm{ph} & (\omega^2_1 + 2 \gamma_1^2) \otimes \hat{\mathds{1}}_\mathrm{ph} & \cdots \\
            \vdots & \vdots & \ddots
        \end{bmatrix}
    \end{gather}
\end{subequations}
where the identity operator for the photonic DOFs, $\hat{\mathds{1}}_\mathrm{ph}$, is explicitly written to emphasize that these are vector/matrices of operators. Now, the Coulomb gauge Hamiltonian can be represented as
\begin{align}
    \hat{H}_\mathrm{p \cdot A} =&~ \hat{H}_\mathrm{M} -  \sqrt{\frac{2}{\hbar}} \bigg( \sum_{ j} \frac{z_j \hat{\bf p}_j}{m_j} \bigg)  \cdot \bigg( \Vec{\hat{\boldsymbol \zeta}}^\intercal \Vec{\hat{q}}_\mathrm{ph} \bigg) \\
    &+ \frac{1}{2} \Big( \Vec{\hat{p}}_\mathrm{ph}^\intercal \Vec{\hat{p}}_\mathrm{ph} +  \Vec{\hat{q}}_\mathrm{ph}^\intercal \tensor{\hat{g}} \Vec{\hat{q}}_\mathrm{ph} \Big). \nonumber
\end{align}
Since $\tensor{\hat{g}}$ is symmetrical and real, it can be diagonalized with an orthogonal matrix, $\tensor{\hat{o}}$. Additionally, $\tensor{\hat{g}}$ is a positive definite matrix. It's eigenvalues are all positive, so its diagonalized form can be written as
\begin{equation}
    \tensor{\hat{o}} \, \tensor{\hat{g}} \, \tensor{\hat{o}}^\intercal = \tensor{\hat{\Omega}} \, \tensor{\hat{\Omega}},
\end{equation}
where $(\tensor{\hat{\Omega}})_{\alpha,\alpha'} = \Omega_\alpha \delta_{\alpha,\alpha'}$ are matrix elements of $\tensor{\hat{\Omega}}$ and $\{ \Omega_\alpha \}$ are the frequencies of the normal modes, $\{ \alpha \}$. As such, the coordinate and momentum operators of the $\alpha_\mathrm{th}$ normal mode are $( \tensor{\hat{o}} \Vec{\hat{q}}_\mathrm{ph} )_\alpha$ and $( \tensor{\hat{o}} \Vec{\hat{p}}_\mathrm{ph} )_\alpha$, respectively.
Additionally, the direction and magnitude of the vector potential of the $\alpha_\mathrm{th}$ normal mode can also be expressed as ${\bf A}_\alpha = (\Vec{\hat{\boldsymbol \zeta}}^\intercal \tensor{\hat{o}}^\intercal)_\alpha / \sqrt{\Omega_\alpha}$.

By expressing the Coulomb gauge Hamiltonian in terms of the $\{ \alpha \}$ normal modes, we recover the Eq.~\ref{eq:HPA-qp} from the main text
\begin{equation}
    \hat{H}_\mathrm{p \cdot A}= \hat{H}_\mathrm{M} -  \sum_{ j,\alpha} \frac{z_j \hat{\bf p}_j \cdot {\bf A}_\alpha}{m_j} \sqrt{\frac{2 \Omega_\alpha}{\hbar}}\hat{q}_\alpha  + \sum_\alpha \frac{1}{2} \Big(\hat{p}_\alpha^2 + \Omega_\alpha^2 \hat{q}_\alpha^2 \Big),
\end{equation}
where we have removed all explicit inter-mode coupling. It should be noted that this normal mode transformation reduces to a Bogoliubov transformation (see Appendix~\ref{app:bogoliubov}) when ${\bf e}_\beta \cdot {\bf e}_{\beta'} = \delta_{\beta, \beta'}$. 

\section{Bogoliubov Transform} \label{app:bogoliubov}
The Bogoliubov transformation~\cite{Bogoljubov1958INC} is a convenient method of partially diagonalizing the additional quadratic terms for Hamiltonians with harmonic oscillators. In the context of the cavity QED Hamiltonian, the normal mode transformation in Appendix~\ref{app:normal_modes} reduces to a Bogoliubov transformation in the single mode and single molecule limit. In this limit, Eq.~\ref{eqn:demlerHc} becomes
\begin{align}\label{eq:HPA-gc-single}
\hat{H}_\mathrm{p \cdot A}=&~ \hat{H}_\mathrm{M} + \hbar \omega_\mathrm{c} (\hat{a}^\dagger \hat{a} + \frac{1}{2}) - \frac{q \hat{\bf p} \cdot {\bf A}_0}{m_j} (\hat{a}^\dagger + \hat{a}) + \frac{\hbar g^2}{2 \omega_\mathrm{c}} (\hat{a}^\dagger + \hat{a})^2,
\end{align}
where $g = |{\bf A}_0| \sqrt{\frac{\omega_\mathrm{c} q^2 }{ \hbar m}}$ is the coupling strength, $\gamma_\mathrm{c}$ in the single-molecule limit.
We then apply this transformation to the following terms of Eq.~\ref{eq:HPA-gc-single}
\begin{equation} \label{eq:demlerHcApp}
\hbar \omega_\mathrm{c} \hat{a}^\dagger \hat{a}  + \frac{\hbar g^2}{2 \omega_\mathrm{c}} (\hat{a}^\dagger + \hat{a})^2.
\end{equation}
To perform this diagonalization, we define new creation and annihilation operators, $\hat{b}^\dagger$ and $\hat{b}$, such that,
\begin{equation}\label{bogo}
    \hat{b} = u \hat{a} + v \hat{a}^\dagger, ~~~\hat{b}^{\dagger} = u \hat{a}^\dagger + v \hat{a},
\end{equation}
where $u$ and $v$ are in real numbers. Requiring the transform to preserve the commutation relation $[\hat{b},\hat{b}^\dagger]=1$ leads to  $[\hat{b},\hat{b}^\dagger]=[u \hat{a} + v \hat{a}^\dagger, u \hat{a}^\dagger + v \hat{a}]=(u^2 - v^2)[\hat{a},\hat{a}^{\dagger}]=1$, thus gives the condition $u^2 - v^2 = 1$. Then, the selection of $(\hat{b}^\dagger + \hat{b})/(\hat{a}^\dagger + \hat{a})$ fully defines the transformation. We want the result of this transformation to be diagonal and enforcing
\begin{equation}\label{bdaggerb}
    \hbar \Omega \hat{b}^\dagger \hat{b} = \hbar \omega_\mathrm{c} \hat{a}^\dagger \hat{a}  + \frac{\hbar g^2}{2 \omega_\mathrm{c}} (\hat{a}^\dagger + \hat{a})^2 - \mathcal{E},
\end{equation}
where $\Omega$ is the dressed photon frequency, and $\mathcal{E}$ is a constant energy shift. Using the transform in Eq.~\ref{bogo}, we expand $\hat{b}^\dagger \hat{b}$ in terms of $\hat{a}$ and $\hat{a}^\dagger$,
\begin{align}
    \hat{b}^\dagger \hat{b} &= (u \hat{a}^\dagger + v \hat{a})(u \hat{a} + v \hat{a}^\dagger)=uv (\hat{a}^{\dagger 2}+\hat{a}^2)+(u^2+v^2)\hat{a}^{\dagger}\hat{a}+v^2\nonumber\\
    &= uv (\hat{a}^\dagger + \hat{a})^2 + (u-v)^2 \hat{a}^\dagger \hat{a} + v^2 - uv,
\end{align}
and comparing to Eq.~\ref{bdaggerb} leads to 
\begin{subequations}
\begin{align}
&(u-v)^2=\omega_\mathrm{c}/\Omega,\label{u-v2}\\ &uv=g^2/2\omega_\mathrm{c}\Omega \label{eq:uv}\\
&\mathcal{E}= \hbar\Omega (uv -v^2).\label{eq:epsilon}
\end{align}
\end{subequations}
Relation Eq.~\ref{u-v2} leads to $u-v=\sqrt{\omega_\mathrm{c}/\Omega}$. This, together with the condition $u^2 - v^2 = (u+v)(u-v)=1$ leads to $u+v=\sqrt{\Omega/\omega_\mathrm{c}}$. Using these two relations, we have,
\begin{equation}\label{uvexpress}
    u = \frac{1}{2} \left[ \sqrt{\frac{\Omega}{\omega_c}} + \sqrt{\frac{\omega_c}{\Omega}} \right], \,\,\,\,\,
    v = \frac{1}{2} \left[ \sqrt{\frac{\Omega}{\omega_c}} - \sqrt{\frac{\omega_c}{\Omega}} \right].
\end{equation}

Note that $\hat{b}^{\dagger}+\hat{b}=(u+v)(\hat{a}^{\dagger}+\hat{a})$, thus the Bogoliubov transformation requires $(\hat{b}^\dagger + \hat{b})= \sqrt{\Omega/\omega_c}(\hat{a}^\dagger + \hat{a})$. Further using Eq.~\ref{eq:uv} and Eq.~\ref{uvexpress}, we have 
\begin{equation}
uv=\frac{1}{4}\left(\frac{\Omega}{\omega_\mathrm{c}}-\frac{\omega_\mathrm{c}}{\Omega}\right)=\frac{g^2}{2\omega_\mathrm{c}\Omega},
\end{equation}
leading to the choice of the frequency 
\begin{equation}\label{norm-freq}
\Omega = \sqrt{\omega_c^2 + 2 g^2}.
\end{equation}
Finally, using Eq.~\ref{eq:epsilon} and Eq.~\ref{uvexpress} we have the expression of the constant $\mathcal{E}=uv-v^2=\frac{\hbar}{2}(\Omega-\omega_c)$, representing the ZPE different associated with two different frequencies. Note that this is positive definite due to $\Omega \ge \omega_c$ and decays to zero as $g \to 0$. Putting all of these together, we can rewrite Eq.~\ref{eqn:demlerHc} in the $\hat{b}$ and $\hat{b}^\dagger$ representation as,
\begin{equation}
    \hat{H}_\mathrm{p \cdot A} = \hat{H}_\mathrm{M} + \hbar \Omega (\hat{b}^\dagger \hat{b} + \frac{1}{2}) -  g \sqrt{\frac{\hbar}{m \Omega}}\mathbf{\hat{e}}\cdot \hat{\mathbf{p}} (\hat{b}^\dagger + \hat{b}).
\end{equation}
This expression is equivalent to Eq.~\ref{eq:shift_pA} in the main text for a single mode.

The basic logic of the Bogoliubov transformation can be understood from a much simpler perspective, with $\hat{q}_\mathrm{c} = \sqrt{\hbar/2\omega_\mathrm{c}}(\hat{a}^{\dagger} + \hat{a})$ and $\hat{p}_\mathrm{c} = i\sqrt{\hbar\omega_\mathrm{c}/2}(\hat{a}^{\dagger} - \hat{a})$ being the photonic coordinate and momentum operators, respectively. Alternatively, 
\begin{subequations}
\begin{align}
\hat{a}^{\dagger}&=\frac{1}{\sqrt{2}}\left(\sqrt{\frac{\omega_\mathrm{c}}{\hbar}}\hat{q}_\mathrm{c}-i\frac{1}{\sqrt{\omega_\mathrm{c}\hbar}}\hat{p}_\mathrm{c}\right)\\
\hat{a}&=\frac{1}{\sqrt{2}}\left(\sqrt{\frac{\omega_\mathrm{c}}{\hbar}}\hat{q}_\mathrm{c}+i\frac{1}{\sqrt{\omega_\mathrm{c}\hbar}}\hat{p}_\mathrm{c}\right).
\end{align}
\end{subequations}
Using $q_\mathrm{c}$ and $p_\mathrm{c}$, one has
\begin{align} 
&\hbar \omega_\mathrm{c} (\hat{a}^\dagger \hat{a}+\frac{1}{2})+ \frac{\hbar g^2}{2 \omega_\mathrm{c}} (\hat{a}^\dagger + \hat{a})^2=\frac{1}{2}\hat{p}^2_\mathrm{c}+\frac{1}{2}\omega^2_{c}\hat{q}^2_\mathrm{c}+g^2\hat{q}^2_\mathrm{c}\nonumber\\
&=\frac{1}{2}\hat{p}^2_\mathrm{c}+\frac{1}{2}(\omega^2_\mathrm{c}+2g^2)\hat{q}^2_\mathrm{c}\equiv\frac{1}{2}\hat{p}^2_\mathrm{c}+\frac{1}{2}\Omega^2\hat{q}^2_\mathrm{c},
\end{align}
where we have introduced $\Omega^2=\omega^2_\mathrm{c}+2g^2$. Introducing the new raising and lowering operators associated with the dressed frequency $\Omega$ as follows
\begin{subequations}
\begin{align}
\hat{b}^{\dagger}&=\frac{1}{\sqrt{2}}\left(\sqrt{\frac{\Omega}{\hbar}}\hat{q}_\mathrm{c}-i\frac{1}{\sqrt{\Omega\hbar}}\hat{p}_\mathrm{c}\right)\\
\hat{b}&=\frac{1}{\sqrt{2}}\left(\sqrt{\frac{\Omega}{\hbar}}\hat{q}_\mathrm{c}+i\frac{1}{\sqrt{\Omega\hbar}}\hat{p}_\mathrm{c}\right), 
\end{align}
\end{subequations}
which naturally gives the condition of Bogoliubov transform $(\hat{b}^\dagger + \hat{b})= \sqrt{\Omega/\omega_c}(\hat{a}^\dagger + \hat{a})$, as well as satisfies Eq.~\ref{bogo} using the coefficients in Eq.~\ref{uvexpress}.

In the case of many interacting particles with mass $m_{j}$ and change $q_{j}$, we start with a more general form of the Coulomb gauge Hamiltonian in Eq.~\ref{eqn:demlerHc},
\begin{align}\label{eqn:demlerHc1}
\hat{H}_\mathrm{p \cdot A} =&~ \hat{H}_\mathrm{M} + \hbar \omega_\mathrm{c} \hat{a}^\dagger \hat{a} 
- \sum_j \frac{z_j \hat{\mathbf{p}}_j \cdot \mathbf{A_0}}{m_j} (\hat{a}^\dagger + \hat{a}) \\
&+ \sum_j \frac{z_j^{2}|\mathbf{A_0} |^2 }{2m_j} (\hat{a}^\dagger + \hat{a})^2, \nonumber
\end{align}
where $j$ is the index of the charged particles. 
The Bogoliubov transformation for this case can also be done, by using the many particle coupling parameter,
\begin{equation}\label{gamma}
    \gamma_\mathrm{c} = |\mathbf{A_0}| \sqrt{ \left( \frac{\omega_c}{\hbar} \right) \sum_j \frac{z_j^2}{m_j}},
\end{equation}
so the terms that the quadratic terms in Eq.~\ref{eqn:demlerHc1} are expressed as
\begin{equation}
\hbar \omega_\mathrm{c} \hat{a}^\dagger \hat{a}  + \frac{\hbar \gamma_\mathrm{c}^2}{2 \omega_\mathrm{c}} (\hat{a}^\dagger + \hat{a})^2.
\end{equation}

Following the same procedure of the Bogoliubov transform is $(\hat{b}^\dagger + \hat{b})= \sqrt{\Omega/\omega_c}(\hat{a}^\dagger + \hat{a})$ with the dressed frequency $\Omega = \sqrt{\omega_c^2 + 2 \gamma_\mathrm{c}^2}$, we now have the Coulomb gauge Hamiltonian as,
\begin{align} \label{eq:HcDemlerMany}
    \hat{H}_\mathrm{p \cdot A} =& \hat{H}_\mathrm{M} + \hbar \Omega (\hat{b}^\dagger \hat{b}+\frac{1}{2})- \sqrt{\frac{\omega_c}{\Omega}} \sum_j \frac{z_j A_0}{m_j} \mathbf{\hat{e}}\cdot \hat{\mathbf{p}}_j (\hat{b}^\dagger + \hat{b}).
\end{align}
As shown in the last line of Eq.~\ref{eq:HcDemlerMany}, the term that is linear in $\hat{\bf{A}}$ is now much more complex. 

For $N$ identical charged particles (such as electrons), the effective coupling parameter in Eq.~\ref{gamma} becomes $\gamma_\mathrm{c}=\sqrt{N}g$, where $g$ is the single electron-cavity coupling strength (Eq.~\ref{eq:g}). 
\begin{align}
    \hat{H}_\mathrm{p \cdot A} =& \hat{H}_\mathrm{M} + \hbar \Omega (\hat{b}^\dagger \hat{b}+\frac{1}{2})- \sqrt{N}g \sqrt{\frac{\omega_c}{\Omega}} \sum_j\mathbf{\hat{e}}\cdot \hat{\mathbf{p}}_j (\hat{b}^\dagger + \hat{b}),
\end{align}
resulting in the familiar collective coupling between light and matter that scales as $\sqrt{N}g$. 

\section{Equivalency to Bloch's Theorem} \label{app:bloch}
From Ref.~[\citenum{AshcroftMermin}], Bloch's Theorem directly follows from the assertion that a periodic potential can be expressed as (in our notation),
\begin{equation} \label{eq:Bloch}
    V(x) = \sum_\kappa v(\kappa) e^{i \kappa x},
\end{equation}
where the $\sum_\kappa$ is over all possible reciprocal lattice vectors, and $v(\kappa)$ can be defined from Eq.~\ref{eq:v_k_expr-3D}. Eq.~\ref{eq:Bloch} and Eq.~\ref{V-pot-conv-3D} from the main text are equivalent as shown by taking the inverse Fourier transform of Eq.~\ref{eq:period_potential-3D}:
\begin{align}
    V(x) &= \int dK \,\, {v}(K) \cdot \Sh (\frac{a_0 K} {2 \pi}) e^{-i K x} \\
    &= \sum_\kappa v(\kappa) e^{-i \kappa x} \nonumber \\
    &= \sum_\kappa v(\kappa) e^{i \kappa x}, \nonumber
\end{align}
where the second line takes advantage of the sifting property of the Dirac delta function and the third line comes from the relation that $v(\kappa) = v(-\kappa)$ by making the assumption that the crystal has inversion symmetry.

We decided to use the convolutional notation in the main text due to the increased flexibility in defining models that it provides. With this method, $v(x)$ can be defined outside of a single unit cell, allowing for models that use functions, such as Gaussians or error functions, that are defined over all space. 

\section{Average Photon Numbers for Double Well Potentials}\label{app:fock_states}
\begin{figure}
 \centering
 %%-------start of first figure----------
    \begin{minipage}[h]{\linewidth}
     \centering
     \includegraphics[width=\linewidth]{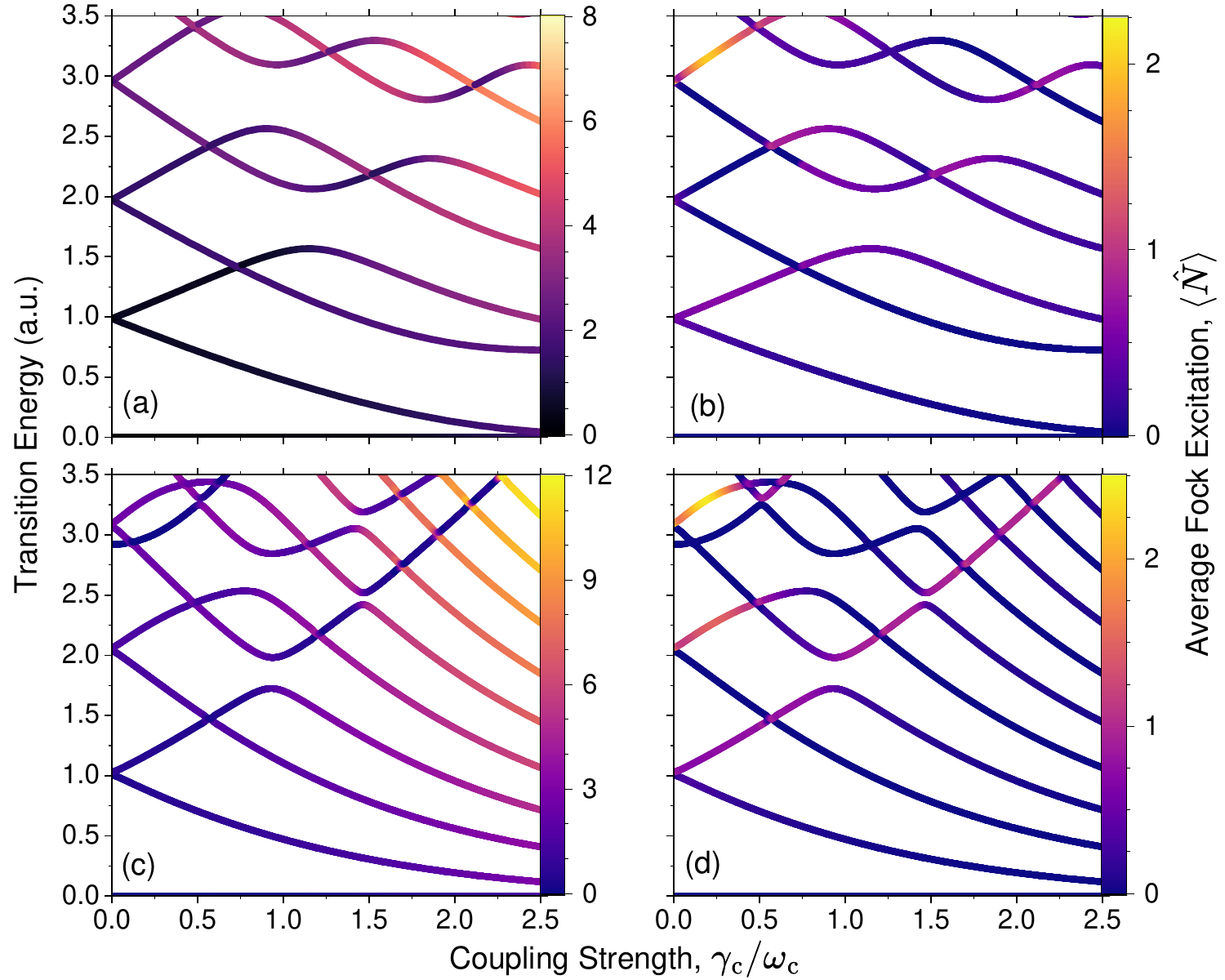}
    \end{minipage}%
   \caption{The polariton states energies as a function of the coupling strength $\gamma / \omega_\mathrm{c}$, color-coded with the average Fock state excitation number. Panels (a)-(b) present the case for the shallow well (see Fig.~\ref{FIG:MAIN_POTS_WFNS_DIPS} for its potential).  (a,b); steep well, (c,d)] for each Hamiltonian [PF, (a,c); RAD, (b,d)] The expectation value of the Fock state excitation number operator is computed in each respective Fock basis, which is not the same between the PF and RAD Hamiltonians but are related by unitary transformation. }
   \label{FIG:MAIN_Photon_Number}
\end{figure}

We provide an analysis of the effective Fock state excitation number as a way to probe the convergence of the RAD Hamiltonian. It is important to note that after unitary transformations, one needs to perform the same transformation on the photonic operator and the quantum states. More detailed discussions can be found in Sec 2.3.4 in Ref.~\cite{Mandal2023CR}. For our current discussion, we are only interested in the efficiency of convergence of the RAD and PF Hamiltonians with Fock states.

Figs.~\ref{FIG:MAIN_Photon_Number}a-d present the eigenspectra of the two double-well model potentials (shown in Fig.~\ref{FIG:MAIN_POTS_WFNS_DIPS}a-b) in a given range of coupling strength up to $\gamma_\mathrm{c}/\omega_\mathrm{c} =2.5$, obtained by diagonalizing (a,c) the PF Hamiltonian and (b,d) the RAD Hamiltonian. The color coding of the curves reflects the value of the average Fock-state excitation number, calculated as follows
\begin{equation}
    N_\psi = \langle \psi | \hat{N} | \psi \rangle,
\end{equation}
where $\hat{N}$ is the Fock state excitation number in a given QED Hamiltonian, and $|\psi \rangle$ is a polariton state. For the PF Hamiltonian, $\hat{N} = \hat{a}^\dag\hat{a}$, and for the RAD Hamiltonian, $\hat{N} = \hat{b}^\dag\hat{b}$. Figs.~\ref{FIG:MAIN_Photon_Number}a-b show results for the steep-well potential (in Fig.~\ref{FIG:MAIN_POTS_WFNS_DIPS}a) for the PF (panel a) and RAD (panel b) Hamiltonians. Note the different color bar scales. PF showcases up to 7 photons (Fock states occupation) on this scale, whereas RAD only shows less than 2. This enables faster convergence in terms of the Fock state basis for the RAD Hamiltonian compared to the PF Hamiltonian. We also note that, at higher coupling strengths, the average Fock state excitation number in the RAD Hamiltonian is much smaller; whereas, for the PF Hamiltonian, the Fock state excitation number increases as the coupling strength increases. This implies that the PF Hamiltonian requires many more Fock basis states to converge the result at any value of coupling compared to the RAD Hamiltonian. Figs.~\ref{FIG:MAIN_Photon_Number}c,d show the same information but for the shallow-well potential (in Fig.~\ref{FIG:MAIN_POTS_WFNS_DIPS}b). One can see that compared to the steep well model potential presented in panels (a)-(b), the PF Hamiltonian requires more Fock states to converge (due to the higher Fock state excitation number),  while the RAD Hamiltonian again shows that its Fock state excitation number goes to zero as the coupling increases for states in the same low-energy range (due to the intrinsic asymptotically decoupled nature of the light-matter interaction in RAD, see Fig.~\ref{FIG:MAIN_Xi_gc__loglog_Demler}a).

\section{Derivation of Pauli-Fierz Hamiltonian} \label{app:PF}
We first introduce the Power-Zienau-Woolley (PZW) gauge transformation operator~\cite{Power1959PTRSA,CohenTannoudji1997} as
\begin{equation}\label{eqn:PZW}
\hat{U}=\exp \big[-\frac{i}{\hbar}\hat{\boldsymbol\mu}\cdot\hat{\bf A}\big]=\exp \big[-\frac{i}{\hbar}\hat{\boldsymbol \mu}\cdot{\bf A}_{0}\big(\hat{a}+\hat{a}^{\dagger}\big)\big],
\end{equation}
or $\hat{U}=\exp \big[-\frac{i}{\hbar}\sqrt{2\omega_\mathrm{c}/\hbar}\hat{\boldsymbol\mu} {\bf A}_0\hat{q}_\mathrm{c}\big]=\exp \big[-\frac{i}{\hbar} (\sum_{j}{z}_j \hat{\bf A} \hat{\bf x}_j)\big]$.
Recall that a momentum boost operator $\hat{U}_\mathrm{p}= e^{-\frac{i}{\hbar} p_{0}\hat{q}}$ displaces $\hat{p}$ by the amount of $p_0$, such that $\hat{U}_\mathrm{p} \hat{O}(\hat{p}) \hat{U}_\mathrm{p}^\dagger = \hat{O}(\hat{p} + p_0)$. Hence, $\hat{U}$ is a boost operator for both the photonic momentum $\hat{p}_\mathrm{c}$ by the amount of $\sqrt{2\omega_\mathrm{c}/\hbar}\hat{\boldsymbol \mu}{\bf A_0}$, as well as for the matter momentum $\hat{\bf p}_{j}$ by the amount of ${z}_j \hat{\bf A}$. 
The PZW gauge operator (Eqn.~\ref{eqn:PZW}) is a special case of $\hat{U}_{\chi}$, such that ${\chi}=-\hat{\bf x}_j\cdot\hat{\bf A}$.
% where ${\boldsymbol\chi}$ now also explicitly dependents on $\hat{\bf A}$ (as appose to a pure function of matter coordinates).  
Using $\hat{U}^{\dagger}$ to boost the matter momentum, one can show that
\begin{equation}\label{boosthm}
\hat{H}_\mathrm{p \cdot A}=\hat{U}^{\dagger}\hat{H}_\mathrm{M}\hat{U}+\hat{H}_\mathrm{ph},
\end{equation}
hence $\hat{H}_\mathrm{p \cdot A}$ can be obtained~\cite{Stefano2019NP} by a momentum boost with the amount of $-{z}_j \hat{\bf A}$ for $\hat{\bf p}_{j}$, then adding $\hat{H}_\mathrm{ph}$. 

 The QED Hamiltonian under the {\it dipole} gauge (the ``$\mathrm{d\cdot E}$'' form~\cite{Power1959PTRSA,GoeppertMayer2009AP}) can be obtained by performing the PZW transformation on $\hat{H}_\mathrm{p \cdot A}$ as follows 
\begin{align}\label{eqn:ddote}
&\hat{H}_\mathrm{d \cdot E}=\hat{U}\hat{H}_\mathrm{p \cdot A}\hat{U}^\dagger=\hat{U} \hat{U}^{\dagger}\hat{H}_\mathrm{M}\hat{U}\hat{U}^{\dagger} +\hat{U}\hat{H}_\mathrm{ph}\hat{U}^\dagger\\
&=\hat{H}_\mathrm{M}+ \hbar \omega_\mathrm{c} (\hat{a}^\dagger \hat{a} + \frac{1}{2}) + i\omega_\mathrm{c} \hat{\boldsymbol \mu} \cdot{\bf A}_0 (\hat{a}^\dagger - \hat{a}) + \frac{\omega_\mathrm{c}}{\hbar}(\hat{\boldsymbol \mu}\cdot{\bf A}_0)^2 \nonumber,
\end{align}
where we have used Eqn.~\ref{boosthm} to express $\hat{H}_\mathrm{p \cdot A}$, and the last three terms of the above equation are the results of $\hat{U}\hat{H}_\mathrm{ph}\hat{U}^\dagger$. Using $\hat{q}_\mathrm{c}$ and $\hat{p}_\mathrm{c}$, one can instead show that 
\begin{equation}
\hat{H}_\mathrm{d \cdot E}=\hat{H}_\mathrm{M}+\frac{1}{2}\omega_\mathrm{c}^{2}\hat{q}_\mathrm{c}^{2}+\frac{1}{2}(\hat{p}_\mathrm{c}+\sqrt{\frac{2\omega_\mathrm{c}}{\hbar}}\hat{\boldsymbol \mu}{\bf A}_0)^2,
\end{equation}
because the PZW operator boosts the photonic momentum $\hat{p}_\mathrm{c}$ by $\sqrt{2\omega_\mathrm{c}/\hbar}\hat{\boldsymbol \mu}{\bf A}_0$. The term $\frac{\omega_\mathrm{c}}{\hbar}(\hat{\boldsymbol \mu}{\bf A}_0)^2$ is commonly referred to as the dipole self-energy (DSE). 

The widely used Pauli-Fierz (PF) QED Hamiltonian~\cite{Flick2017PNAS,Schaefer2018PRA,Rokaj2018JPBAMOP} in recent studies of polariton chemistry can be obtained by using the following unitary transformation 
\begin{equation}
\hat{U}_{\phi}=\exp[-i\frac{\pi}{2}\hat{a}^{\dagger}\hat{a}].
\end{equation}
Note that $\hat{U}_{\phi}\hat{a}^{\dagger}\hat{a}\hat{U}^{\dagger}_{\phi}=\hat{a}^{\dagger}\hat{a}$, $\hat{U}_{\phi} \hat{a}\hat{U}^{\dagger}_{\phi} = i\hat{a}$, and $\hat{U}_{\phi}\hat{a}^{\dagger} \hat{U}^{\dagger}_{\phi} = -i\hat{a}^{\dagger}$, applying $\hat{U}_{\phi}$ on $\hat{H}_\mathrm{d \cdot E}$, we have the PF Hamiltonian as follows
\begin{align}\label{eqn:pf_1}
&\hat{H}_\mathrm{PF}=\hat{U}_{\phi}\hat{H}_\mathrm{d \cdot E}\hat{U}^{\dagger}_{\phi}\\
&=\hat{H}_\mathrm{M}+ \hbar \omega_\mathrm{c} (\hat{a}^\dagger \hat{a} + \frac{1}{2}) + \omega_\mathrm{c} \hat{\boldsymbol\mu}\cdot{\bf \bf A}_0( \hat{a}+\hat{a}^\dagger) + \frac{\omega_\mathrm{c}}{\hbar}(\hat{\boldsymbol\mu}\cdot{\bf A_0})^2\nonumber\\
&=\hat{H}_\mathrm{M}+\frac{1}{2}\hat{p}_\mathrm{c}^2 + \frac{1}{2}\omega_\mathrm{c}^2\big(\hat{q}_\mathrm{c} + \sqrt{\frac{2}{\hbar\omega_\mathrm{c}}}\hat{\boldsymbol\mu}\cdot{\bf A_0}\big)^2\nonumber
\end{align}
The above PF Hamiltonian has the advantage of a pure real Hamiltonian and the photonic DOF can be viewed~\cite{Flick2017PNAS,Schaefer2018PRA} and computationally treated~\cite{Hoffmann2019JCP,Li2020JCP} as an additional ``nuclear coordinate''. 

\begin{figure}[]
 \centering
 %%-------start of first figure----------
    \begin{minipage}[]{\linewidth}
     \centering
     \includegraphics[width=1.0\linewidth]{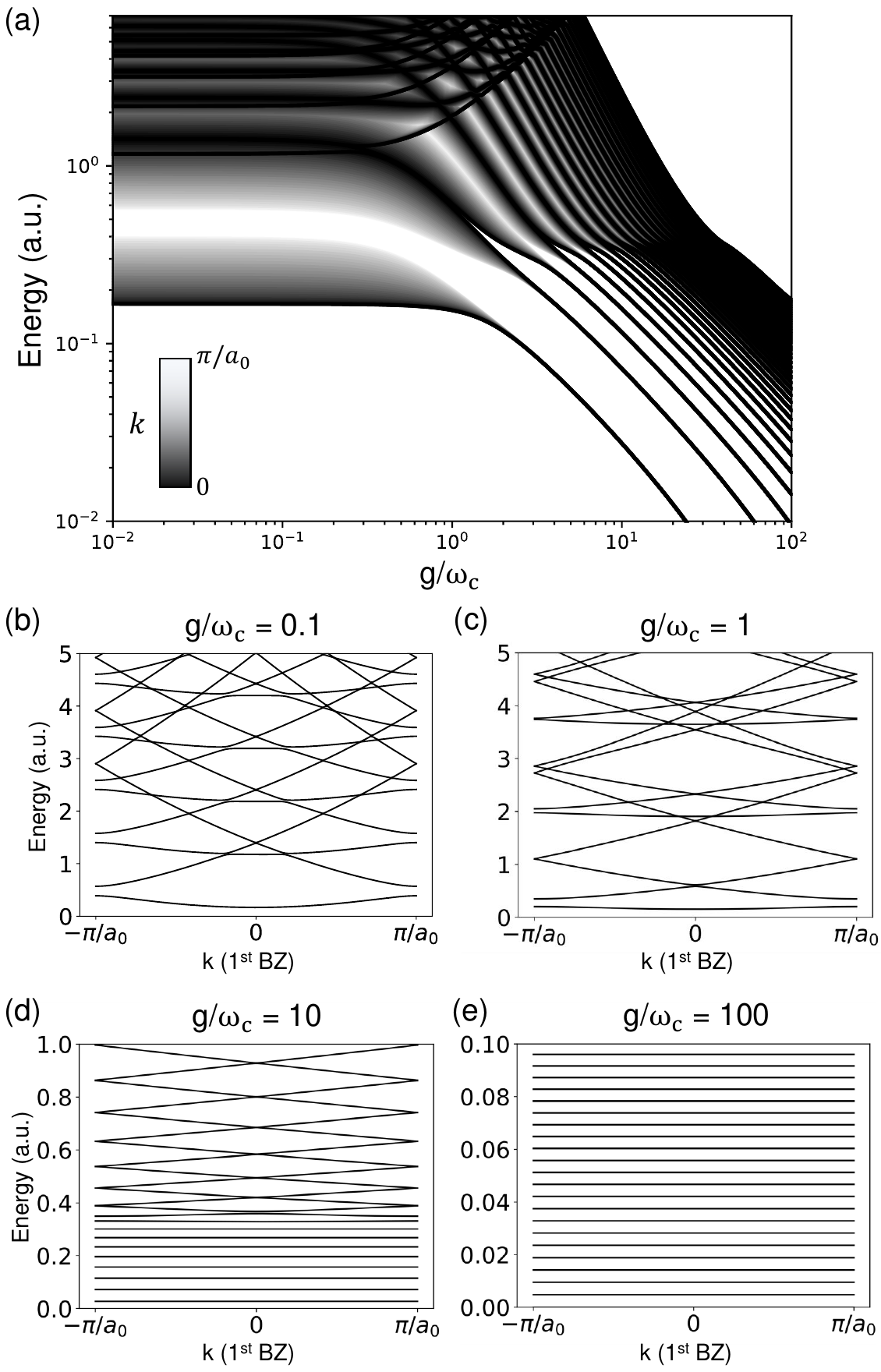}
    \end{minipage}%
   \caption{Single electron in a cosine potential coupled to a cavity. (a) The first 40 bands in the energy eigenspectrum of this model as a function of the normalized coupling strength $g/\omega_\mathrm{c}$ color-coded by the $k$-point. (b)-(e) Dispersion plots for normalized coupling strengths of $g/\omega_\mathrm{c} = 0.1, 1, 10, 100$, respectively. }
   \label{fig:cos_pot}
\end{figure}
\section{Details of Numerical Calculations} \label{app:DVR}
The model matter Hamiltonians were diagonalized using the discrete variable representation (DVR) \cite{colbert_novel_1992} for the electron kinetic energy $\hat{T}$ such that the Hamiltonian in the position basis for a uniform grid takes the form,
\begin{equation}\label{EQ:DVR_HAM}
    H_{xx'} = T_{xx'} + V(x)\delta_{xx'} ,
\end{equation}
\noindent with,
\begin{equation}
    T_{xx'} = \frac{\hbar^2 (-1)^{x-x'}}{2 m (\Delta x)^2 }
\begin{Bmatrix}
\frac{\pi^2}{3}, & x = x'\\
 & \\
\frac{2}{(x-x')^2}, & x \ne x'
\end{Bmatrix},
\end{equation}
\noindent where $\Delta x$ is the grid spacing with $\hbar^2/ 2 m \Delta x^2$ as the energy quantum of the grid and $x$ and $x'$ as the real-space grid indices. This Hamiltonian is diagonalized directly using standard approaches for Hermitian matrices. Additionally, the full light-matter Hamiltonians are diagonalized in the same manner.

To solve the PF Hamiltonian, the electric dipole matrix elements are required, as they mediate the interactions between light and matter. They are calculated from the electronic wavefunctions (eigenstates of Eq.~\ref{EQ:DVR_HAM}) in the usual way as,
\begin{equation}
    %\mu_{\psi \phi} = -|e|~\int~dx~\psi_\psi^*(x)~x~\psi_\phi(x),
    \mu_{\psi \phi} = -|e|~\int~dx~\langle\psi|x\rangle x \langle x|\phi\rangle,
\end{equation}
\noindent where $e$ is the elementary electric charge and $|\phi\rangle$ is the $\phi_\mathrm{th}$ electronic wavefunction of Eq.~\ref{EQ:DVR_HAM}. All dipole matrix elements were solved with 2048 grid points to converge the lowest 50 electronic states and all transition dipole moments between them.

The Fourier transform of the real-space potential $V(x) \rightarrow V(k)$ in each model was computed using the asymmetrically normalized forward-backward fast Fourier transform (FFT) implemented in Python (with the normalization defined in Eq.~\ref{eq:FT}) by the NumPy module without padding.

\section{Application on 1D Cosine Potential}
We further provide additional examples of using RAD Hamiltonian to solve 1D periodic potential. We define a cosine potential as,
\begin{align}
    V(x) =&~ v_0 \cos (k_0 x) \\
    \mathcal{V}(K) =&~ \frac{v_0}{2} \left( \delta(K-k_0) + \delta(K + k_0) \right) \nonumber
\end{align}
where $v_0$ is the amplitude of the cosine and $k_0 = 2 \pi / a_0$. Using the expression of $\hat{{H}}_\mathrm{RAD}$ from Eq.~\ref{eq:RAD_Ham} along with the Fourier Transform of this potential, we can define the Hamiltonian analytically for this model as,
\begin{align} \label{eq:cos_RAD_Ham}
    &\hat{H}_\mathrm{RAD} =  \, \hbar \Omega \hat{b}^\dagger \hat{b} + \int dK' \, \frac{(\hbar K')^2}{2 m_\mathrm{eff}} \ketbra{K'}{K'} \\
    &+ \int dK' \, \frac{v_0}{2} \, \left( \ketbra{K'}{K' + k_0}  \, e^{-i k_0 {\xi_g} \hat{q}_\mathrm{c}} \, + \, \ketbra{K'}{K' - k_0}  \, e^{i k_0 {\xi_g} \hat{q}_\mathrm{c}} \right). \nonumber
\end{align}

Furthermore, we can then find the dispersion plots using an $\hat{{H}}_\mathrm{RAD}(k)$ of the form,
\begin{align} \label{eq:cos_h_rad_k}
    \hat{H}_\mathrm{RAD}(k) =&  \, \hbar \Omega \hat{b}^\dagger \hat{b} \otimes \hat{P}_k + \sum_{\kappa}  \, \frac{\hbar^2 ( k +\kappa)^2}{2 m_\mathrm{eff}} \ketbra{k +\kappa}{k +\kappa} \\
    &+ \sum_{\kappa} \, \frac{v_0}{2} \, \Big( \ketbra{k +\kappa}{k +\kappa + k_0}  \, e^{-i k_0 {\xi_g} \hat{q}_\mathrm{c}} \nonumber \\
    &+ \, \ketbra{k +\kappa}{k +\kappa - k_0}  \, e^{i k_0 {\xi_g} \hat{q}_\mathrm{c}}\Big) , \nonumber
\end{align}
This is clearly a special case of Eq.~\ref{eq:h_rad_k_gen}, where the only non-zero off-diagonal terms in the Hamiltonian occur when $\kappa' = \pm 2\pi / a_0$. 

Figure \ref{fig:cos_pot} shows the numerical results of this cosine model. Fig.~\ref{fig:cos_pot}(a) shows how the eigenspectrum changes as the coupling strengths evolve through the ultrastrong coupling regime and into the deep-strong coupling regime. Panels (b)-(d) show the dispersion plots of the polariton states for four different coupling strengths. 

%\nocite{*}
\bibliographystyle{unsrt}
%\bibliography{references.bib}% Produces the bibliography via BibTeX.

\end{document}